\documentclass[twocolumn]{aastex62}   

\usepackage{comment}

\usepackage{natbib}
\usepackage{multirow}
\usepackage{color}
\usepackage{bm}
\usepackage[normalem]{ulem}
 \usepackage{amsmath} 
\usepackage {diagbox}
\usepackage{ltablex}
\usepackage{longtable}
\usepackage{threeparttable}

\usepackage{mathtools}

\newcommand{\Nifs}{$^{56}$Ni}

\def\gsim{\mathrel{\rlap{\lower 4pt \hbox{\hskip 1pt $\sim$}}\raise 1pt \hbox {$>$}}}
\def\lsim{\mathrel{\rlap{\lower 4pt \hbox{\hskip 1pt $\sim$}}\raise 1pt \hbox {$<$}}}

\newcommand{\ltsim}{\protect\raisebox{-0.8ex}{$\:\stackrel{\textstyle <}{\sim}\:$}}
\newcommand{\gtsim}{\protect\raisebox{-0.8ex}{$\:\stackrel{\textstyle >}{\sim}\:$}}

\shorttitle{W-shaped O II lines in SLSNe-I}
\shortauthors{S.Saito et al.}

\begin{document}

\title{On the Formation of the W-shaped O II Lines in Spectra of Type I Superluminous Supernovae}

\correspondingauthor{Sei Saito}
\email{s.saito@astr.tohoku.ac.jp}

\author{Sei Saito}
\affiliation{Astronomical Institute, Tohoku University, Aoba, Sendai 980-8578, Japan}

\author{Masaomi Tanaka}
\affiliation{Astronomical Institute, Tohoku University, Aoba, Sendai 980-8578, Japan}
\affiliation{Division for the Establishment of Frontier Sciences, Organization for Advanced Studies, Tohoku University, Sendai 980-8577, Japan}

\author{Paolo A. Mazzali}
\affiliation{Astrophysics Research Institute, Liverpool John Moores, University, IC2, Liverpool Science Park, 146 Brownlow Hill, Liverpool L3 5RF, UK}
\author{Stephan Hachinger}
\affiliation{Leibniz Supercomputing Centre (LRZ) of the Bavarian Academy of Sciences and Humanities, Boltzmannstr. 1, 85748 Garching b.M., Germany}

\author{Kenta Hotokezaka}
\affiliation{Research Center for the Early Universe, Graduate School of Science, University of Tokyo, Bunkyo, Tokyo 113-0033, Japan}

\begin{abstract}

H-poor superluminous supernovae (SLSNe-I) are characterized
by O II lines around  $4,000  - 4,500 ~ {\rm \AA}$ in pre-/near-maximum spectra, 
so-called W-shaped O II lines.
As these lines are from relatively high excitation levels,
they have been considered a sign of non-thermal processes,
which may give a hint of power sources of SLSNe-I.
However, the conditions for these lines to appear have not been understood well.
In this work, we systematically calculate synthetic spectra to reproduce observed spectra of eight SLSNe-I,
parameterizing departure coefficients from the nebular approximation in the SN ejecta (expressed as $b_{\rm neb}$).
We find that most of the observed spectra can be reproduced well with $b_{\rm neb} \ltsim 10$,
which means that no strong departure is necessary for the formation of the W-shaped O II lines.
We also show
that the appearance of the W-shaped O II lines is sensitive to temperature;
only spectra with temperatures $T \sim 14,000 - 16,000$ K can produce the W-shaped O II lines without large departures.
Based on this,
we constrain the non-thermal ionization rate near the photosphere.
Our results suggest that spectral features of SLSNe-I can give independent constraints on the power source through the non-thermal ionization rates.

 \end{abstract}


\section{Introduction}
\label{sec:int}

Superluminous supernovae (SLSNe) are known as a special type of supernovae (SNe) with extremely high luminosities.
SNe with their peak absolute magnitudes of \ltsim $- 21$ mag in optical bands are conventionally classified as SLSNe \citep[e.g.,][]{Gal-Yam2012, Moriya2018a, Gal-Yam2019aug, Nicholl2021}.
Also, some SNe less luminous than $\sim - 21$ mag at their peak magnitudes are often classified as SLSNe by the similarity of the spectra to SLSNe
\citep[e.g.][]{Decia2018, Quimby2018, Gal-Yam2019}.

Power sources of SLSNe are still under debate.
Several candidate power sources have been proposed:
decay of a large amount of \Nifs ~ \citep[as synthesized by pair-instability SNe, ][]{Barkat1967, Heger2002},
interaction with circumstellar material \citep[CSM, ][]{Smith2007, Chevalier2011, Ginzburg2012, Chatzopoulos2012}, 
and central engines such as magnetars \citep{Ostriker1971, Kasen2010, Woosley2010}
or black hole accretion \citep{Dexter2013, Moriya2018}.

SLSNe are divided into two subgroups based on their spectroscopic features:
Type I SLSNe (SLSNe-I), which do not show H lines in their spectra \citep[e.g.,][]{Quimby2011} and
Type II SLSNe (SLSNe-II), which do \citep[e.g.,][]{Smith2010}.
Spectra of SLSNe-I tend to show characteristic W-shaped absorption lines around $4,000  - 4,500 {\ \rm \AA}$ in pre-/near-maximum phases.
The O II lines are thought to give the dominant contribution to these features, and thus,
these features are called the W-shaped O II lines
\citep{Chomiuk2011, Quimby2011, Lunnan2013, Nicholl2015, Branch2017, Liu2017, Gal-Yam2019, Kumar2020, Konyves-Toth2021, Konyves-Toth2022}
These features are not seen in normal SNe except for a few possible cases:
Type Ib SN 2008D \citep{Modjaz2009},
Type Ibn OGLE-2012-SN-006 \citep{Pastorello2015}, and
Type IIL SN 2019hcc \citep{Parrag2021}.
Also, some SLSNe-I, such as SN 2015bn, do not show the W-shaped O II lines \citep{Konyves-Toth2021}.

\begin{figure}[t]
\centering
\includegraphics[width= 0.98\linewidth, bb= -10 -10 1050 800]{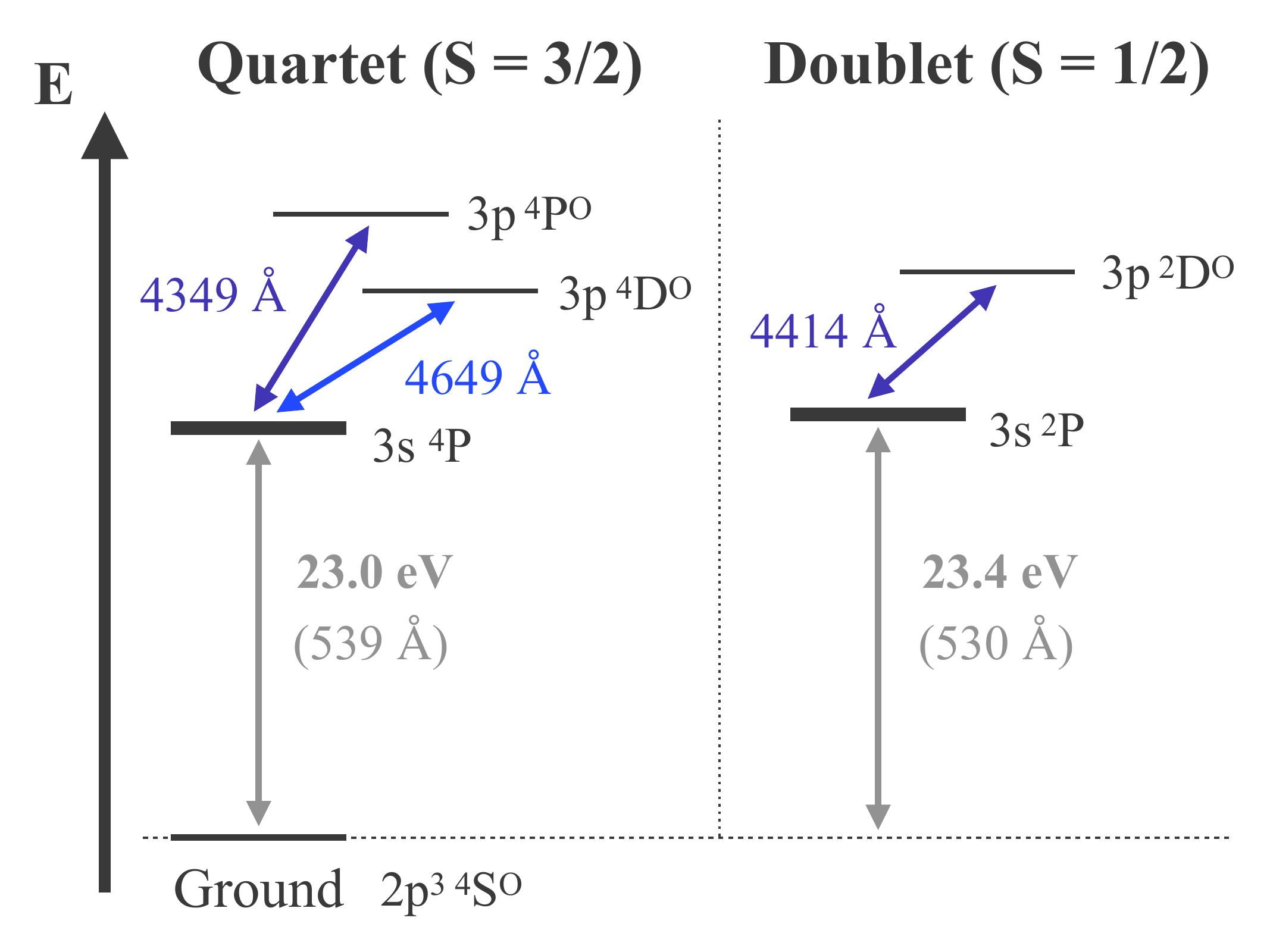}
\caption{
A schematic energy diagram of O II mainly contributing the W-shaped O II lines.
}
\label{fig:config}
\end{figure}

The W-shaped O II lines are mainly produced by bound-bound transitions
from the 3s $^4$P (the spin-quartet state) and 3s $^2$P (the spin-doublet state) levels.
The lower energy levels of these lines are $\sim 23$ eV above the ground state (Figure \ref{fig:config}).
If the distribution across the excited states follows the Boltzmann distribution in local thermodynamic equilibrium (LTE),
such high energy levels require high temperatures to be populated so that strong absorption lines can be produced.
Therefore, non-thermal processes (in ionization and recombination and excitation) causing departure from LTE
have been considered necessary to produce the W-shaped O II lines \citep[e.g., ][]{Mazzali2016, Quimby2018}.
However, it is not yet clear whether non-thermal processes are required.
In fact, \citet{Dessart2019} suggested that the W-shaped O II lines can appear
without the effect of non-thermal processes by performing radiative transfer simulations.

Departure from LTE is reminiscent of He I lines in spectra of He-rich SNe (SNe Ib).
The prominent absorption lines of He I in SN Ib spectra arise from excited states
with energies $\sim 20$ eV above the ground state.
Since this energy is much higher than the typical temperature of SNe Ib ($T \sim 5,000 - 10,000$ K or $\sim 0.4 \ -  0.9$ eV),
the energy levels are not populated enough to produce the absorption lines in LTE.
This suggests that they are populated by non-thermal processes.
The departure from LTE can be understood well by $\gamma-$rays from \Nifs ~ decay \citep{Lucy1991, Hachinger2012}.
The excited states of He I are populated by high energy electrons produced by Compton scattering of $\gamma$-rays from \Nifs \ decay.
The high energy electrons ionize He I to He II, and then He II recombines to the excited states of He I \citep[see Figure 1 of][]{Tarumi2023}.
\citet{Lucy1991} suggested that the degree of departure from LTE described by the departure coefficient $b$
is $\sim 10^4 - 10^5$ for the He I lines in spectra of SNe Ib
($n = b \times n^{\ast}$, where $n$ and $n^{\ast}$ are the number density of atoms in the excited state in ejecta of actual SNe Ib and in LTE, respectively).

As the departure from LTE for the He I lines provides information on the power source of SNe Ib,
the departure from LTE to produce the W-shaped O II lines in spectra of SLSNe-I may give us a hint of the power source of SLSNe-I.
However, the detailed physical conditions where these lines appear are not understood well.
In this work, we investigate the conditions for the appearance of the W-shaped O II lines
by calculating synthetic spectra and comparing them with observed spectra.
Then, we demonstrate that spectra of SLSNe-I can be used to give constraints on the power source.

This paper is organized as follows.
Section \ref{sec:obs} describes our samples of observed spectra of SLSNe-I.
Section \ref{sec:calc} gives an overview of the methods used for our calculations and parameters to calculate synthetic spectra.
Section \ref{sec:res} shows results of the spectral calculations and comparison with the observed data.
Section \ref{sec:dis} discusses the behavior of the departure coefficients and implications on the power source of SLSNe-I.
Finally, Section \ref{sec:sum} summarizes the findings of this paper.

\section{Observational data}
\label{sec:obs}

\begin{figure*}[th]
\centering
\includegraphics[width= 0.7 \linewidth]{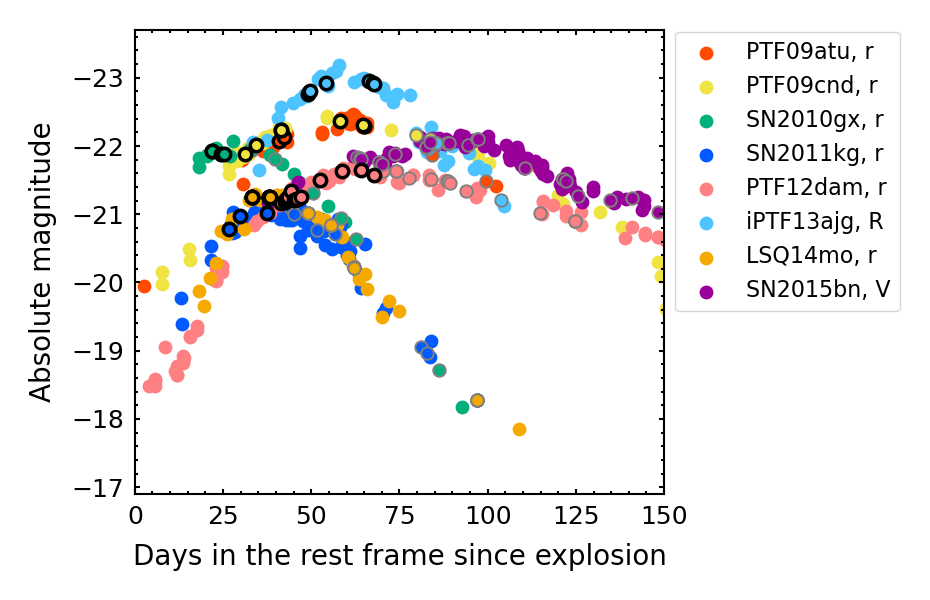}
\caption{
Light curves of the eight SLSNe-I in our sample.
We show light curves in the bands with the best temporal coverage.
The difference in the colors correspond to the different SLSNe-I as shown in the legend.
The points show the photometric observations.
The points with the black edges show epochs when spectra with      the W-shaped O II lines have been taken.
The points with the gray edges  show epochs when spectra without the W-shaped O II lines have been taken.
}
\label{fig:lc}
\end{figure*}

We use 66 spectra of eight SLSNe-I (see Table \ref{table:sample}) taken from the
WISeREP database \citep{Yaron2012} and corresponding photometric data from the Open
Supernova Catalog \citep[][light curves are shown in Figure
\ref{fig:lc}]{Guillochon2017jan}. The spectra were selected from spectra of the 28
SLSNe-I analyzed in \citet{Konyves-Toth2021} based on the following two criteria: 
i) availability of photometric data within three days of the spectra  
and
ii) availability of spectroscopic or photometric data at least at one epoch
covering wavelengths shorter than $\lambda_{\rm rest} \ltsim$ 2,000 $\rm \AA$ in 
the rest frame.
The former condition enables accurate flux calibration of the spectra
while the latter one secures an accurate determination of temperatures of SLSNe-I (typically $T \sim 10,000 - 15,000$ K).
Accurate temperature estimates are necessary to evaluate the departure coefficient for the appearance of the W-shaped O II lines
since the LTE population strongly depends on temperature in the Boltzmann factor.

For flux calibration, we scaled the fluxes of the observed spectra to the corresponding photometry.
To obtain a photometric flux at each epoch when a spectrum was taken,
we linearly interpolated two photometric data points in the magnitude space.
Wavelength dependence in the flux scaling factor was considered as a linear function of wavelength
if a spectrum has two or more corresponding photometric data.
If a spectrum has only one corresponding photometric data,
the whole spectrum was multiplied by a constant scaling factor.

Milky Way extinction was corrected for based on the dust map by \citet{Schlegel1998}.
Note that extinction within the host galaxy was not corrected for
since its amount is often uncertain.
We will discuss possible inaccuracies in Section \ref{dis:13ajg}.
When the observed spectra are compared with synthetic spectra (described in Section \ref{sec:calc}),
the distance modulus of each SLSN-I is applied to the synthetic spectra.
As SNe in our sample are at moderate redshifts (Table \ref{table:sample}),
distance moduli are obtained from the redshifts with the cosmological parameters
$H_0 = 70 ~ \rm{km ~ s^{-1} ~ Mpc^{-1}}, \Omega_{\rm m} = 0.3, and ~ \Omega_{\lambda} = 0.7$.

\begin{table}[t]
\centering
\begin{threeparttable}[flushleft]
  \caption{Sample of SLSNe-I}
  \label{table:sample}
  \begin{tabular}{lccc}
    \hline \hline
    Name &  $z$ & $E(B-V)$ & $t_{\rm rise}$\tnote{1}  \\
         &      & mag      &  days   \\    
    \hline
PTF09atu    &  0.5015  & 0.0409   &  59  \\ 
PTF09cnd   & 0.2584  & 0.0207     &  57   \\ 
SN 2010gx  &  0.2299  & 0.0333    & 29   \\ 
SN 2011kg  &  0.1924  & 0.0371    & 33    \\ 
PTF12dam  &  0.1074  & 0.0107     &  64    \\ 
iPTF13ajg   &  0.7400  & 0.0121   &  59    \\ 
LSQ14mo   &  0.2530  & 0.0646     &  40    \\ 
SN 2015bn & 0.1136  & 0.0221      &  91   \\ 
\hline 
  \end{tabular}
  \begin{tablenotes}
   \raggedright
      \item[1] The rise time in the bands shown in Figure \ref{fig:lc}.
  \end{tablenotes}
\end{threeparttable}
\end{table}

\section{Spectral calculations}
\label{sec:calc}

\subsection{Code}

For calculations of synthetic spectra, we utilize a one-dimensional Monte Carlo radiative transfer code \citep{Mazzali1993, Mazzali2000}.
Here, we briefly describe the code by
focusing on the assumptions relevant to this work.
The code calculates the plasma conditions under the
modified nebular approximation.
The modified nebular approximation takes into account the fact that,
in the SN atmosphere,
the densities are low and radiative processes dominate,
giving the largest influence on the level population
\citep{Abbott1985,Mazzali1993}.
Thus, the code does not assume LTE both for ionization and excitation.
The modified nebular approximation is known to work well for spectra of normal SNe \citep[e.g.,][]{Pauldrach1996, Sauer2006, Tanaka2008, Hachinger2012, Teffs2020}.
For more details of the code, we refer readers to \citet{Mazzali1993} and \citet{Mazzali2000}.

The code assumes a sharp photosphere inside the ejecta.
At the photosphere, blackbody radiation is emitted.
Photon packets replicating flux are then propagated through the SN ejecta outside the photosphere.
For the photon propagation in the ejecta,
electron scatterings and bound-bound absorptions are taken into account.
The degree of scattering and absorption is determined by the level populations and 
ionization in the ejecta, which depend on density and temperature.
A temperature structure in the ejecta outside the photosphere is established by tracing the photon packets,
differentiating between a local radiation temperature ($T_{\rm R}$) and an electron temperature ($T_{\rm e}$).
The local electron temperature is crudely assumed to be $T_{\rm e} = 0.9 \ T_{\rm R}$.
The temperatures are estimated by an iterative process
as the temperatures give level populations and ionization degrees,
which in turn affect the energy flux, and thus the temperatures.

The level population (number density) of the $j$-th ionized element with atomic 
number $i$ at the $l$-th excited level $n_{i, j, l}$ is (at some given radius):
\begin{equation}
\label{eq:excite}
\frac{n_{i, j, l}}{n_{i, j, 0}} = W \frac{g_{i, j, l}}{g_{i, j, 0}} e^{-E_{i, j, l}/kT_{\rm R}},
\end{equation}
where
$g_{i, j, l}$ is the statistical weight,
$E_{i, j, l}$ is the excitation energy from the ground state,
$k$ is the Boltzmann constant,
and 
$T_{\rm R}$ is a radiation temperature.
$W$ in Equation (\ref{eq:excite}) is the dilution factor calculated as 
\begin{equation}
\label{eq:j}
J = W B(T_{\rm R}),
\end{equation}
where
$J$ is the frequency-integrated mean intensity from the simulation and
$B(T_{\rm R})$ is the frequency-integrated Planck function with the radiation temperature $T_{\rm R}$ at some given radius.

Ionization is estimated by the modified nebular approximation \citep{Mazzali1993}:
\begin{equation}
\label{eq:ionize}
\frac{n_{i, j+1, 0} n_{\rm e }}{n_{i, j, 0}}
=
\eta W \sqrt{\frac{T_{\rm e}}{T_{\rm R}}}
2 \frac{g_{i, j+1,0}}{g_{i,j,0}}
\frac{(2 \pi m_{\rm e} k T_{\rm R})^{3/2}}
{h^3} e^{-I_{i, j}/kT_{\rm R}},
\end{equation}
where 
$n_{\rm e}$ is the number density of electrons,
$m_{\rm e}$ is the mass of the electron,
$h$ is the Planck constant,
and $I_{i, j}$ is the ionization potential of the $j$-th ionized element with 
atomic number $i$.
In Equation (\ref{eq:ionize}) $\eta$ is defined as 
\begin{equation}
\eta = \delta \zeta + W(1-\zeta),
\end{equation}
where
$\delta$ is a correction factor for the optically thick region at wavelengths shorter than $\lambda_0 = 1,050 \ {\rm \AA}$ (the Ca II edge),
and $\zeta$ is the fraction of recombinations going directly to the ground state.
Typically, $\eta \sim 0.4$ for O just outside the photosphere ($W \sim 0.5$).

Line scattering by bound-bound transitions from a lower level $l$ to an upper level $u$
in homologously expanding ejecta is treated in the Sobolev approximation \citep{Sobolev1957}.
A particularly important quality in this context is the Sobolev optical depth:
\begin{align}
\label{eq:tau}
\tau_{lu}
&= \frac{hc}{4 \pi}(B_{lu}n_{i,j,l} - B_{ul}n_{i,j,u})t_{\rm expl} \nonumber \\
&\simeq \frac{\pi e^2}{m_{\rm e} c} f_{lu} n_{i, j, l} t_{\rm expl} \lambda_{lu},
\end{align}
where
$B_{lu}$ and $B_{ul}$ are Einstein B-coefficients,
$f_{lu}$ is the oscillator strength of the transition,
$t_{\rm expl}$ is the time since explosion, 
and $\lambda_{lu}$ is the wavelength of the transition.
The effect of line fluorescence is also taken into account \citep{Mazzali2000}.

\subsection{Set up of calculations} 
\label{sec:setup}

The density structure of the ejecta outside the photosphere follows a power law 
($\rho \propto r^{-n}$, where $\rho$ is density and $r$ is radius) 
with index $n=7$.
The radius can be expressed by a velocity $v$ in homologously expanding ejecta
($r = vt_{\rm expl}$).
In our fiducial model (iPTF13ajg, see Section \ref{sec:res}), 
the ejecta mass above $v = 12,000 {\rm \ km \ s^{-1}}$
is $M_{\rm ej} (v \geq 12,000 {\rm \ km \ s^{-1}}) = 3.7 \ M_\odot$.
This ejecta mass is scaled by $f_{\rho}$ parameter as described below.
Note that the total ejecta mass cannot be estimated from our modeling
as the radiative transfer is solved only outside of the photosphere.
Thus, in the following sections, we give the mass outside of a typical velocity,
for which we adopt $v = 12,000 {\rm \ km \ s^{-1}}$, as a representative value.
Abundances in the ejecta are assumed to be homogeneous,
and set to be the same as those adopted by 
\citet{Mazzali2016} for the first spectrum of iPTF13ajg 
for all the 66 spectra analyzed in this work:
$X({\rm He}) = 0.10$, $X({\rm C}) = 0.40$, $X({\rm O}) = 0.475$, $X({\rm Ne}) = 0.02$, 
$X({\rm Mg}) = 5 \times 10^{-4}$, $X({\rm Si}) = 2 \times 10^{-3}$, $X({\rm S}) = 5 \times 10^{-5}$, 
$X({\rm Ca}) = 5 \times 10^{-5}$, $X({\rm Ti}) = 2 \times 10^{-5}$, $X({\rm Fe}) = 5 \times 10^{-4}$, 
$X({\rm Co}) = 3 \times 10^{-4}$, and $X({\rm Ni}) = 1 \times 10^{-4}$.

Parameters of the spectral calculations are as follows:
 \begin{itemize}
  \item $f_{\rho}$: density scaling factor
  \item $L_{\rm bol}$: bolometric luminosity  [${\rm erg \ s^{-1}}$]
  \item $v_{\rm ph}$: velocity at the photosphere [${\rm km \ s^{-1}}$]
  \item $t_{\rm expl}$: time since explosion in the rest frame [${\rm days}$]
  \item $b_{\rm neb}$: departure coefficient for the excited states of O II.
 \end{itemize}
Note that the purpose of this work is to understand the condition for the W-shaped O II lines as compared with those of normal SNe.
Thus, we define the departure coefficient $b_{\rm neb}$ as a departure from the population expected from the modified nebular approximation
($n_{\rm O II} = b_{\rm neb} \times n_{\rm O II, neb}$, where $n_{\rm O II}$ and $n_{\rm OII, neb}$ are
the number density of O II in the excited state in ejecta of SLSNe-I
and that in the modified nebular approximation, respectively).

These parameters in our calculations to model the observed spectra can almost independently be determined from the following physical quantities.
We constrain the density scaling parameter $f_{\rho}$ from the dilution factor
so that the dilution factor $W$ at $v_{\rm ph}$ estimated by Equation (\ref{eq:j}) is converged to $\sim 1/2$
(i.e., close to the value of the geometric dilution factor)
for a given combination of the other parameters.
$f_{\rho}$ is kept the same for the spectral series of each SN.
$L_{\rm bol}$ is constrained from the observed flux (with an accuracy of $\sim 10 \%$).
$v_{\rm ph}$ is estimated from wavelengths of blueshifted absorption lines.
$t_{\rm expl}$ is then constrained using $L_{\rm bol} = 4 \pi (v_{\rm ph} t_{\rm expl})^2 \sigma {T_{\rm eff}}^4$,
where $\sigma$ is the Stephan-Boltzmann constant and $T_{\rm eff}$ is the effective temperature.
We estimate $t_{\rm expl}$ for each spectrum so that the entire spectral series for each SN are reproduced by the same explosion date.
The estimated explosion date gives a rise time to the peak of the light curve (Table \ref{table:sample}).
The light curves in Figure \ref{fig:lc} are shown as a function of the time since the estimated explosion date,
and our estimated rise time is consistent with (or longer than)
that estimated from an extrapolation of the light curve.
Finally, $b_{\rm neb}$ is estimated from depth of the W-shaped O II lines.

\section{Results}
\label{sec:res}

\subsection{Synthetic spectra}

\begin{figure}[t]
\centering
\includegraphics[width= 0.98\linewidth]{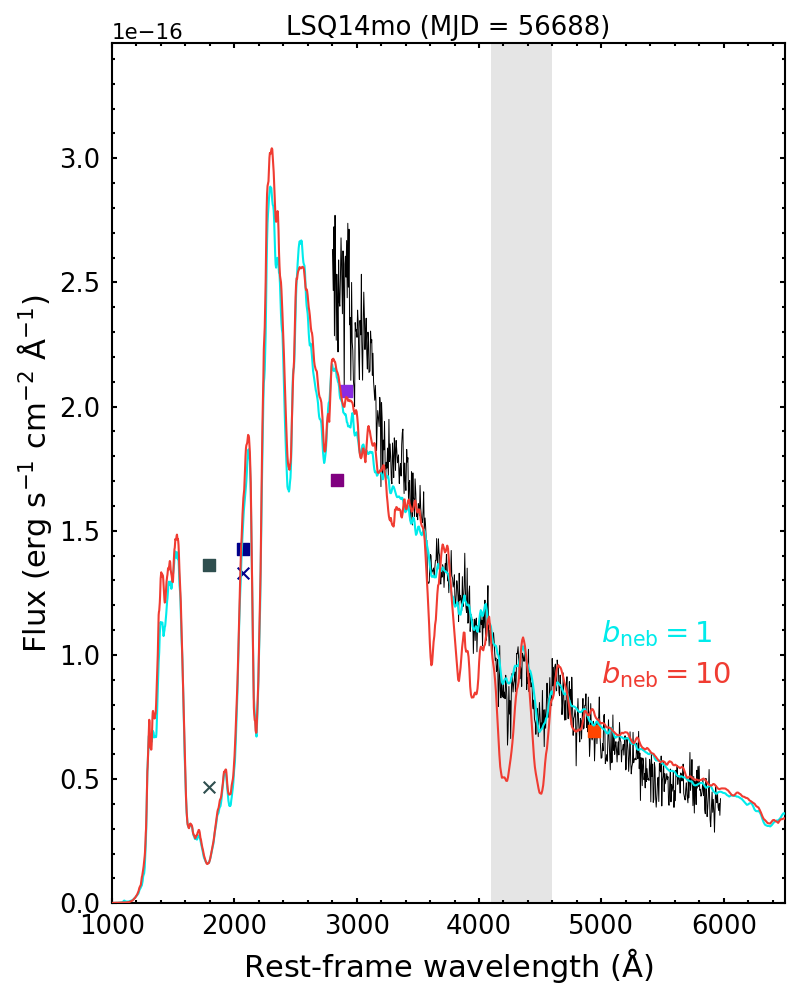}
\caption{
An example of modeling for the first spectrum of LSQ14mo (MJD $=$ 56688).
The black line shows the observed spectrum.
The light blue line shows a synthetic spectrum with the parameters:
$L_{\rm bol} = 10^{44.2} \ {\rm erg \ s^{-1}}$,
$v_{\rm ph} = 10,500 {\rm \ km \ s^{-1}}$, 
$t_{\rm expl} = 33$ days,
$b_{\rm neb} = 1$ (no departure), 
and $f_\rho = 0.2$.
The estimated radiation temperature just outside the photosphere is $T_{\rm R} 
\sim 15,000$ K.
The red line is a synthetic spectrum with the same parameters as those of the 
light blue line but with departure coefficient $b_{\rm neb} = 10$.
The square points show photometric fluxes.
Only the photometric point near 5,000 ${\rm \AA}$ (the $r$-band) was used for flux calibration of the observed spectrum.
The photometric points near 3,000 ${\rm \AA}$ were not used because the coverage 
of the observed spectrum was not wide enough to be compared with the photometry.
The crosses are integrated fluxes of the synthetic spectrum ($b_{\rm neb} = 1$) with the filters of UV where there is no coverage of the observed spectrum.
}
\label{fig:example}
\end{figure}

Figure \ref{fig:example} shows an example of comparison between synthetic spectra and an observed spectrum (the first spectrum of LSQ14mo).
The observed spectrum is reproduced reasonably well with a departure coefficient $b_{\rm neb} = 1$ (no departure from the nebular approximation)
at an estimated radiation temperature just outside the photosphere $T_{\rm R} \sim 15,000$ K.
The model with $b_{\rm neb} = 10$ produces too deep W-shaped O II lines.
The wavelengths of the W-shaped O II lines are matched well with a photospheric velocity $v_{\rm ph} = 10,500 {\rm ~ km \ s^{-1}}$.
The UV fluxes are strongly affected by metal absorption, which makes it difficult to estimate temperatures
when observed spectral energy distributions are simply fitted by blackbody functions.
Models with the best parameters for all the spectra in our sample are shown in Figure \ref{fig:all}
and the best parameters are summarized in Table \ref{table:param} (Appendix \ref{app:param}).
The properties of each SN are descried below.

\begin{figure*}[htbp]
    \begin{tabular}{cc}
      \begin{minipage}[t]{0.5\hsize}
        \centering
        \includegraphics[width= 0.98\linewidth]{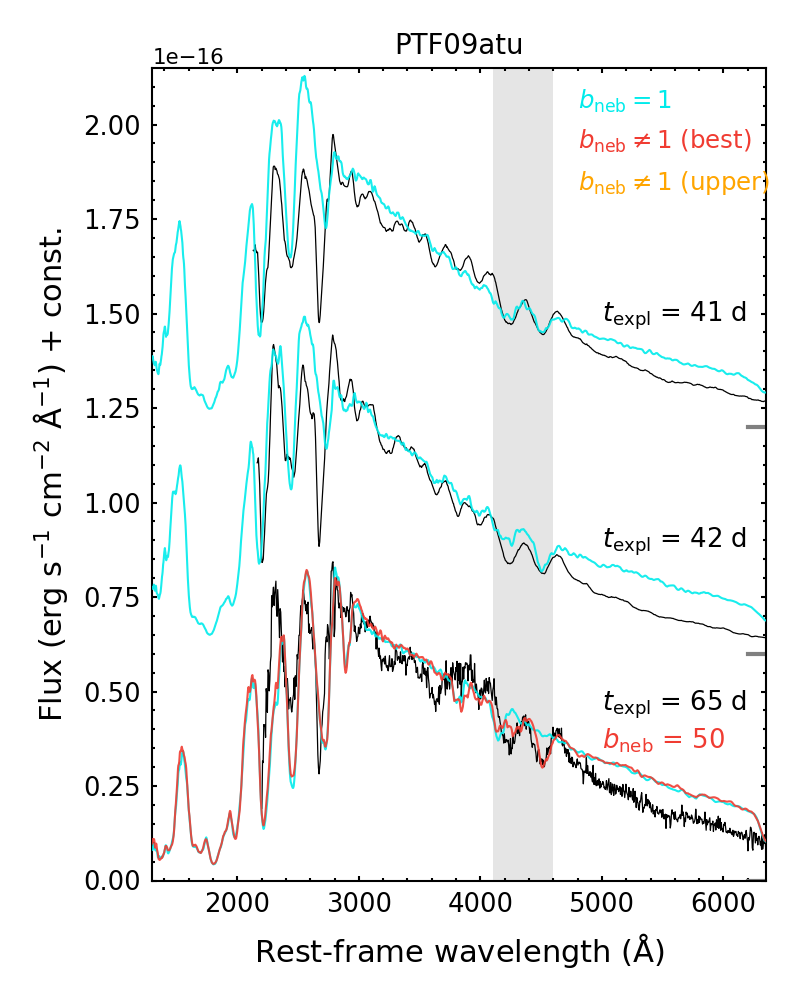}
      \end{minipage} &
      \begin{minipage}[t]{0.5\hsize}
        \centering
        \includegraphics[width= 0.98\linewidth]{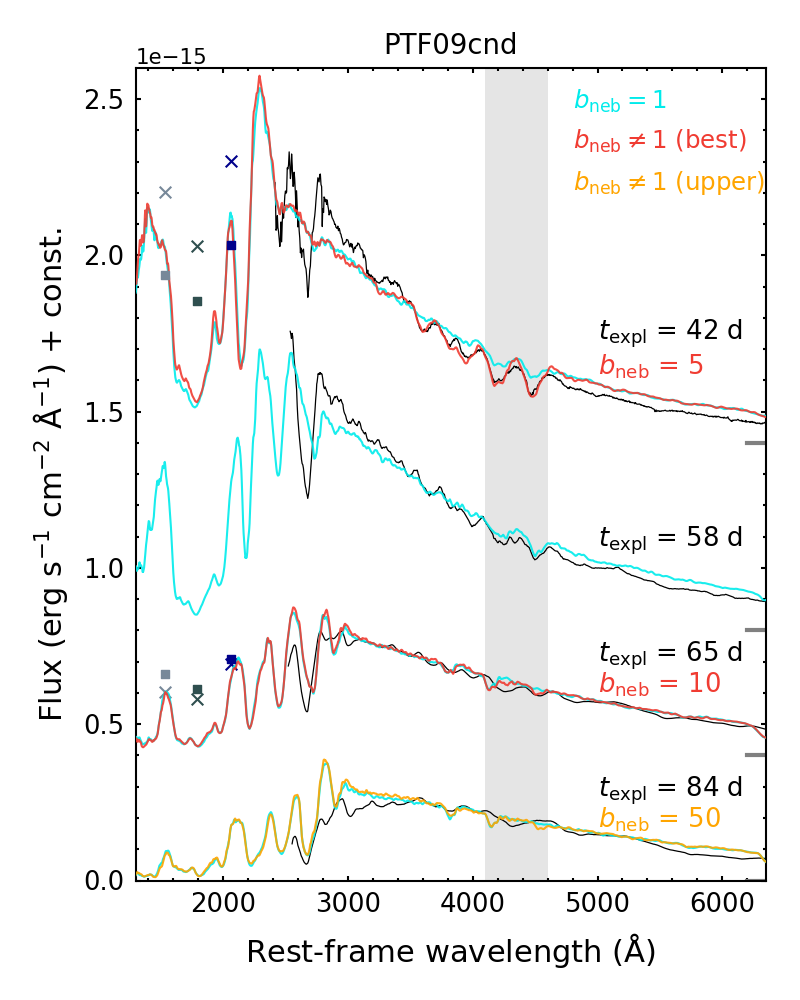}
      \end{minipage} \\
   
      \begin{minipage}[t]{0.5\hsize}
        \centering
        \includegraphics[width= 0.98\linewidth]{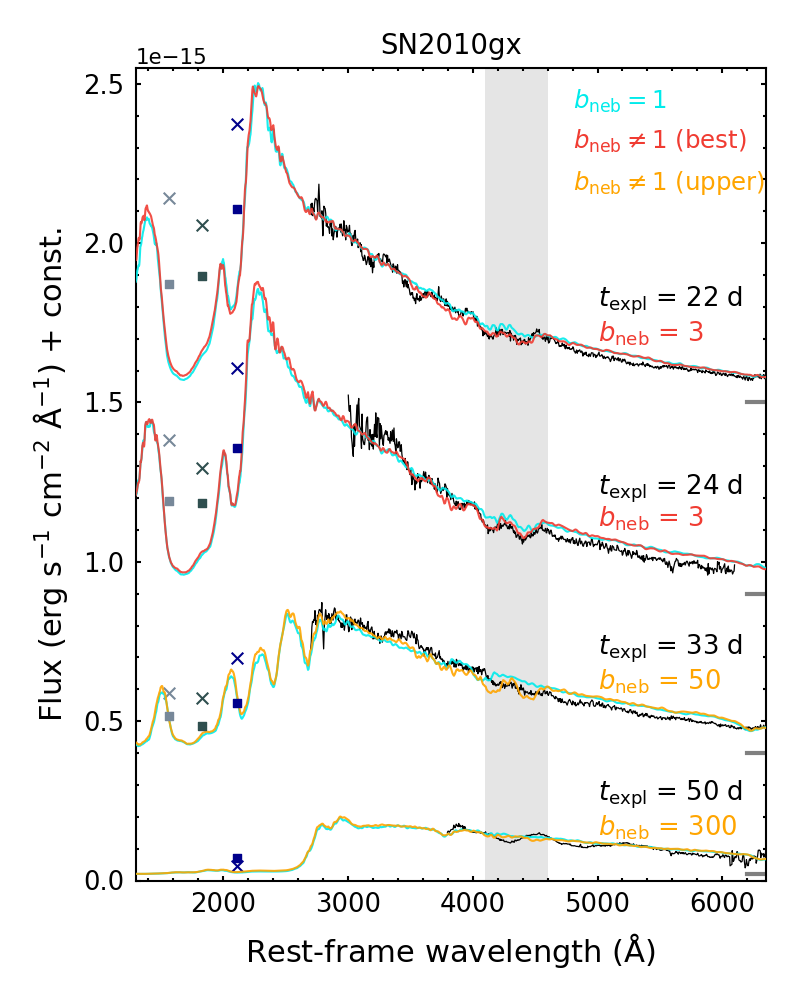}
      \end{minipage} &
      \begin{minipage}[t]{0.5\hsize}
        \centering
        \includegraphics[width= 0.98\linewidth]{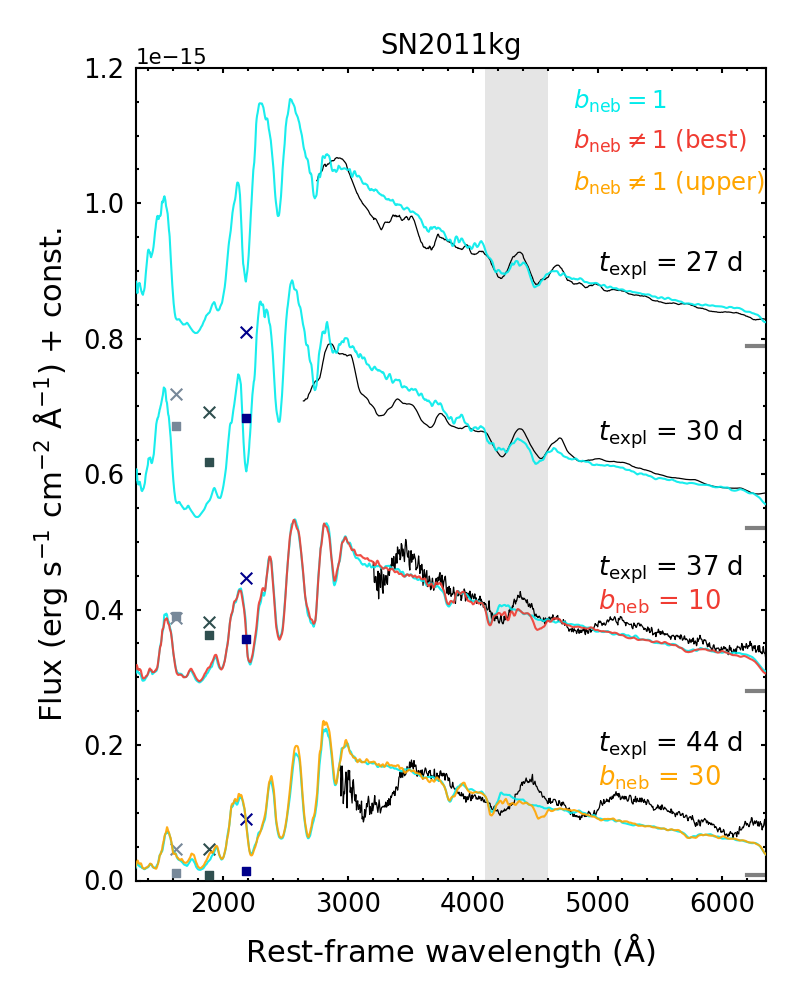}
      \end{minipage} 
    \end{tabular}
     \caption{All the results of modeling. Note that some spectra with other spectra taken at close epochs were excluded for visibility.
For the observed spectra with the W-shaped O II lines reproduced well with $b_{\rm neb}=1$,
we only show one model with $b_{\rm neb}=1$ (the light blue lines) for each observed spectrum.
For the observed spectra with the W-shaped O II lines reproduced well with $b_{\rm neb} > 1$,
we show two models:
a model with $b_{\rm neb} = 1$ (the light blue lines) and a model with $b_{\rm neb} > 1$  (the red lines).
For the observed spectra without the W-shaped O II lines,
we also show two models:
a model with $b_{\rm neb} = 1$ (the light blue lines) and a model with $b_{\rm neb} > 1$  (the orange lines) as an upper limit of $b_{\rm neb}$.
The square points show UV photometric fluxes.
The crosses show integrated fluxes of the synthetic spectra ($b_{\rm neb} = 1$) with the filters of UV.
Vertical offsets for each spectrum are shown by the horizontal lines on the right y-axis.
}
\label{fig:all}
  \end{figure*}

  \begin{figure*}[htbp]
  \addtocounter{figure}{-1}
    \begin{tabular}{cc}
      \begin{minipage}[t]{0.5\hsize}
        \centering
        \includegraphics[width= 0.98\linewidth]{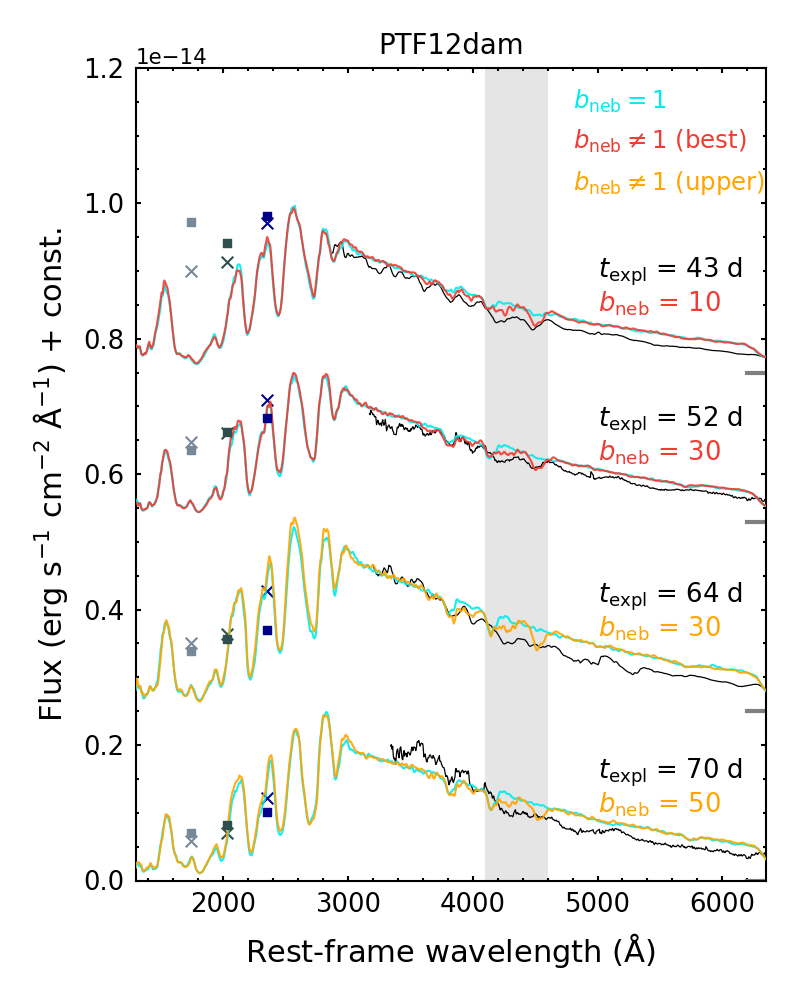}
      \end{minipage} &
      \begin{minipage}[t]{0.5\hsize}
        \centering
        \includegraphics[width= 0.98\linewidth]{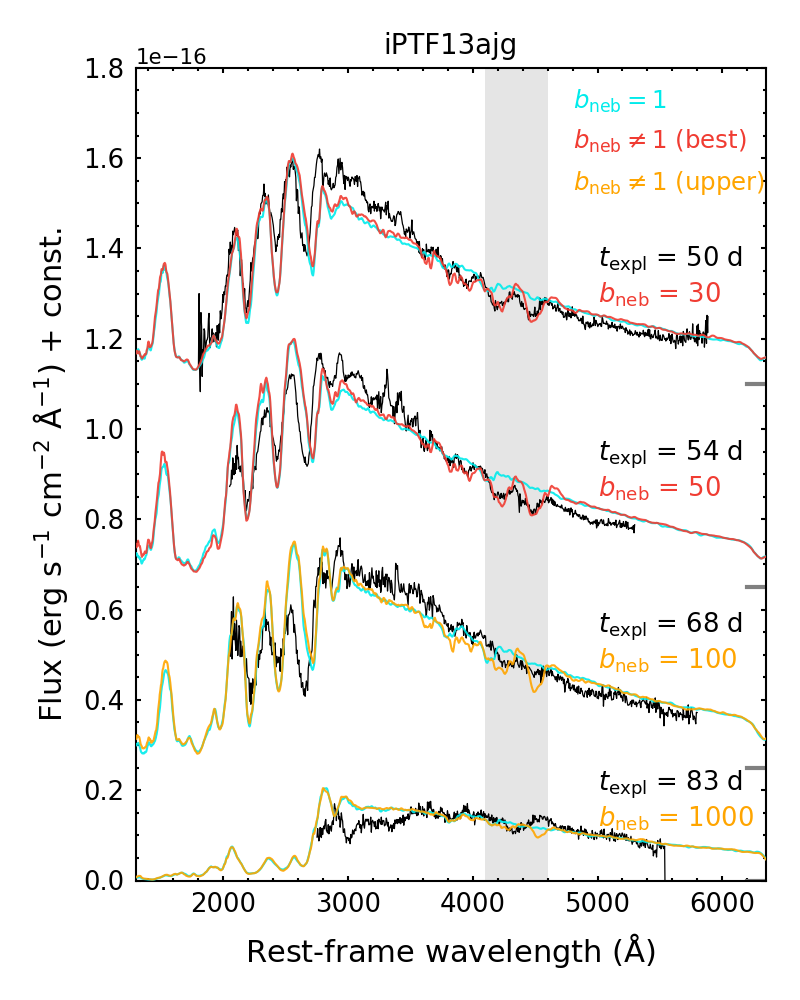}
      \end{minipage} \\
   
      \begin{minipage}[t]{0.5\hsize}
        \centering
        \includegraphics[width= 0.98\linewidth]{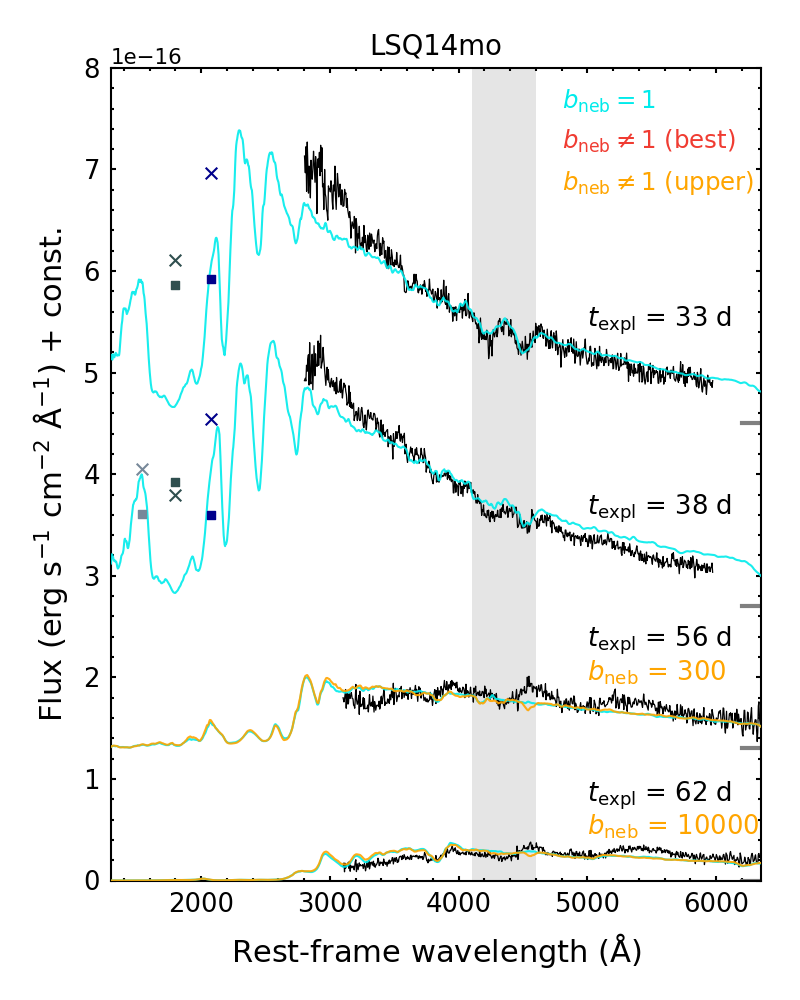}
      \end{minipage} &
      \begin{minipage}[t]{0.5\hsize}
        \centering
        \includegraphics[width= 0.98\linewidth]{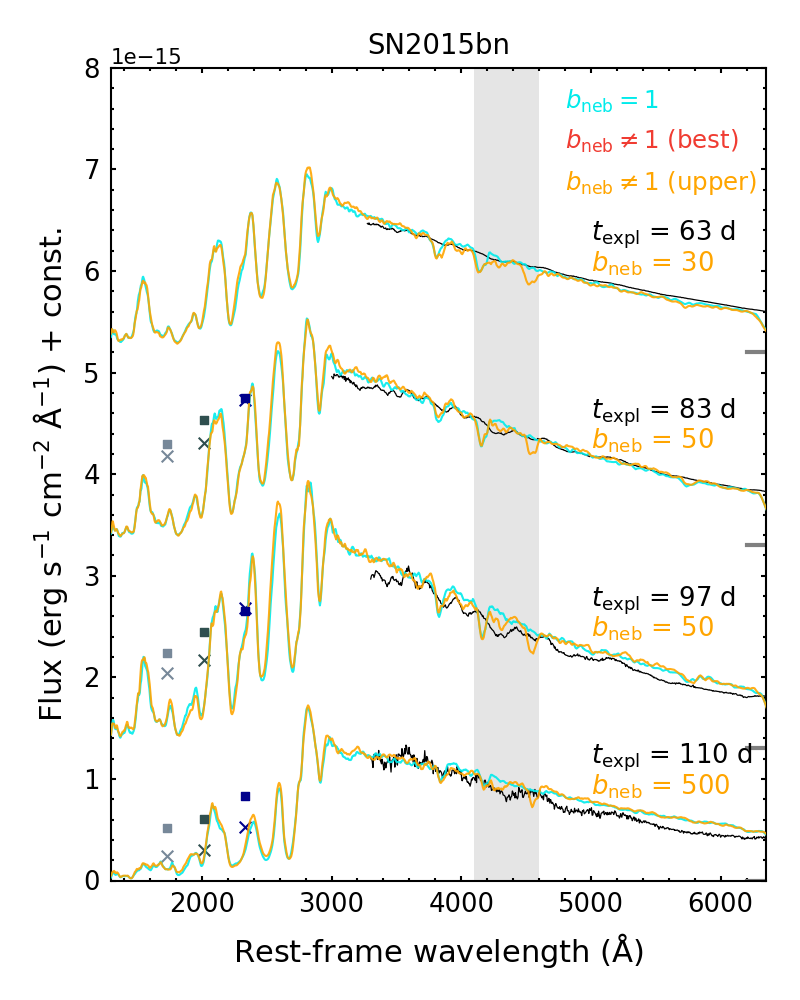}
      \end{minipage} 
    \end{tabular}
     \caption{\textit{Continued.}}
  \end{figure*}

\textbf{iPTF13ajg}:
First, we show results of modeling for iPTF13ajg, which we use as our fiducial model as this object was also modeled in \citet{Mazzali2016}.
This object was reported in \citet{Vreeswijk2014}.
The absolute AB magnitude at $u$-band peak is $\sim -22.5$ mag \citep{Vreeswijk2014}.

For the modeling, we adopt the same parameters as in \citet{Mazzali2016}.
By definition, the density scaling factor is $f_\rho = 1.0$,
which corresponds to the ejecta mass outside above $v = 12,000 \ {\rm km \ s^{-1}}$ of $M_{\rm ej}(v \geq 12,000 \ {\rm km \ s^{-1}}) = 3.7 \ M_\odot$.
This object shows the W-shaped O II lines until $t_{\rm expl} = 54$ days.
The required departure coefficients to reproduce the depths of the W-shaped O II lines are $b_{\rm neb} \sim 50$.
The depths of the W-shaped O II lines of this object require one of the largest departure coefficients among all the objects in our sample.

\textbf{PTF09atu}:
This object was reported in \citet{Quimby2011}.
The absolute AB magnitude at $u$-band peak is $\sim -21.5$ mag \citep{Quimby2011}.
To reproduce the time series of the spectra,
the density scaling factor for this object is found to be $f_\rho = 0.5$,
which corresponds to the ejecta mass outside $v = 12,000 \ {\rm km \ s^{-1}}$
of $M_{\rm ej}(v \geq 12,000 {\rm  \ km \ s^{-1}}) = 1.8 \ M_\odot$.
This object shows the W-shaped O II lines at all the epochs at which spectra were taken ($t_{\rm expl} = 23 - 46$ days).
The required departure coefficients to reproduce the depths of the W-shaped O II lines are $b_{\rm neb} \sim 1 - 50$.

\textbf{PTF09cnd}:
This object was also reported in \citet{Quimby2011}.
The absolute AB magnitude at $u$-band peak is $\sim -22$ mag \citep{Quimby2011}.
The spectra taken at MJD 55055 (2009 August 12) and MJD 55068 (2009 August 25) were also modeled in \citet{Mazzali2016}.
We updated flux calibration of the spectra using the newly published photometric data \citep{Decia2018};
the fluxes are found to be higher than those in \citet{Mazzali2016} by a factor of $\sim 2.5$.
Thus, the bolometric luminosity of the best model in this work is also higher than \citet{Mazzali2016} by a factor of $\sim 2.5$.
With the increase of the bolometric luminosity, the time since explosion is longer than that in \citet{Mazzali2016}
to have a larger photospheric radius.
This updated time is also consistent with the light curve.
The density scaling factor for this object is found to be $f_\rho = 0.7$,
which corresponds to the ejecta mass outside $v = 12,000 \ {\rm km \ s^{-1}}$
of $M_{\rm ej}(v \geq 12,000 {\rm \ km \ s^{-1}}) = 2.6 \ M_\odot$.
Despited the differences in the luminosity and time,
the overall properties of the model are consistent with those derived by \citet{Mazzali2016},
e.g., the inferred density scaling factor (or mass outside of a certain velocity) for PTF09cnd is smaller than that for iPTF13ajg.

This object shows the W-shaped O II lines until $t_{\rm expl} = 65$ days.
The required departure coefficients to reproduce the depths of the W-shaped O II lines are  $b_{\rm neb} \sim 1 - 10$.
Note that the absorption line around 2,700 ${\rm \AA}$ cannot be reproduced
by our models because of the fixed homogeneous abundance. \citet{Mazzali2016}
reproduces those lines by Mg II with the abundance $X$(Mg) $\sim$ 0.1, but we
fixed the abundance to $X$(Mg) $= 5 \times 10^{-4}$.

\textbf{SN 2010gx}:
This object was reported in \citet{Mahabal2010}, \citet{Pastorello2010a}, and \citet{Quimby2011} as PTF10cwr
and in \citet{Pastorello2010} as SN 2010gx.
The absolute AB magnitude at $u$-band peak is $\sim -21.7$ mag \citep{Quimby2011}.
For this object, the density scaling parameter is found to be $f_\rho = 3.0$,
which corresponds to the ejecta mass outside $v = 12,000 \ {\rm km \ s^{-1}}$
of $M_{\rm ej}(v \geq 12,000 {\rm \ km \ s^{-1}}) = 11 \ M_\odot$.
This object shows the W-shaped O II lines until $t_{\rm expl} = 25$ days.
The required departure coefficients to reproduce the depths of the W-shaped O II lines are $b_{\rm neb} \sim 1 - 3$.
After $t_{\rm expl} = 25$ days, the W-shaped O II lines disappear.
The observed spectra taken after MJD 55323 ($t_{\rm expl} = 63$ days) could not be reproduced
maybe because of the fixed abundance and/or the single power density law and the mass assumption.
Thus, we only show models for the observed spectra before MJD 55323.

\textbf{SN 2011kg}:
This object was reported in \citet{Quimby2011tel}.
The absolute AB magnitude at $g$-band peak is $\sim -20.8$ mag \citep{Inserra2013}.
For this object, the density scaling parameter is found to be $f_\rho = 0.1$,
which corresponds to the ejecta mass outside $v = 12,000 \ {\rm km \ s^{-1}}$
of $M_{\rm ej}(v \geq 12,000 {\rm \ km \ s^{-1}}) = 0.4 \ M_\odot$.
This object shows the W-shaped O II lines until $t_{\rm expl} = 37$ days.
The required departure coefficients to reproduce the depths of the W-shaped O II lines are $b_{\rm neb} \sim 1 - 10$ (small departure).
After $t_{\rm expl} = 37$ days, the W-shaped O II lines disappear.
The absorption lines around $\sim 3,000 {\rm \ \AA}$ and $\sim 3,500 {\rm \ \AA}$ 
may be due to Fe II and/or Si II,
which are not reproduced with our fixed abundance.
However, this does not affect the departure coefficient required for the O II lines.

\textbf{PTF12dam}:
This object was reported in \citet{Quimby2012tel}.
The absolute AB magnitude at $g$-band peak is $\sim -22$ mag \citep{Chen2017}.
For this object, the density scaling parameter is found to be $f_\rho = 0.4$,
which corresponds to the ejecta mass outside $v = 12,000 \ {\rm km \ s^{-1}}$
of $M_{\rm ej}(v \geq 12,000 {\rm \ km \ s^{-1}}) = 1.5 \ M_\odot$.
This object shows the W-shaped O II lines until $t_{\rm expl} = 59$ days.
The required departure coefficients to reproduce the depths of the W-shaped O II lines are up to $b_{\rm neb} \sim 50$.
After $t_{\rm expl} = 59$ days, the W-shaped O II lines disappear.

\textbf{LSQ14mo}:
This object was reported in \citet{Nicholl2015}.
The absolute AB magnitude at $g$-band peak is $\sim -21.2$ mag \citep{Leloudas2015}.
The spectra of this object were also modeled in \citet{Chen2017}.
The first spectrum (at $-7$ days) is modeled with $t_{\rm expl} = 18$ days.
However, the model spectrum shows a somewhat higher velocity of the W-shaped O II lines as compared with the observed spectrum. 
To have a better match in the velocity,
we adopt the photospheric velocity smaller than that estimated in \citet{Chen2017} by $\sim 5,000 {\rm \ km \ s^{-1}}$.
Accordingly, the time since explosion in our model for this spectrum is estimated to be $t_{\rm expl} = $ 33 days.
This revised time is compatible with the light curve, and reproduces the later spectra as well.
With this choice,
the density scaling parameter for this object is found to be $f_\rho = 0.2$,
which corresponds to the ejecta mass outside $v = 12,000 \ {\rm km \ s^{-1}}$
of $M_{\rm ej}(v \geq 12,000 {\rm \ km \ s^{-1}}) = 0.7 \ M_\odot$.

This object shows the W-shaped O II lines until $t_{\rm expl} = 54$ days.
The required departure coefficients to reproduce the depths of the W-shaped O II lines are $b_{\rm neb} \sim 1$ (no departure).
After $t_{\rm expl} = 54$ days, the W-shaped O II lines disappear.

\textbf{SN 2015bn}:
This object was reported in \citet{Nicholl2016}.
The absolute AB magnitude at $g$-band peak is $\sim - 22$ mag \citep{Nicholl2016}.
For this object, the density scaling parameter is found to be $f_\rho = 0.4$,
which corresponds to the ejecta mass outside $v = 12,000 \ {\rm km \ s^{-1}}$
of $M_{\rm ej}(v \geq 12,000 {\rm \ km \ s^{-1}}) = 1.5 \ M_\odot$.
This object lacks the W-shaped O II lines throughout the epochs of available spectra ($t_{\rm expl} = 63 - 135$ days).
This is the only object that does not show the W-shaped O II lines at any epochs in our sample.

\subsection{Temperature and departure coefficient}

\begin{figure*}[ht]
\begin{center}
  \begin{tabular}{c}
    \begin{minipage}{0.5\hsize}
      \begin{center}
       \includegraphics[width= 0.97\linewidth]{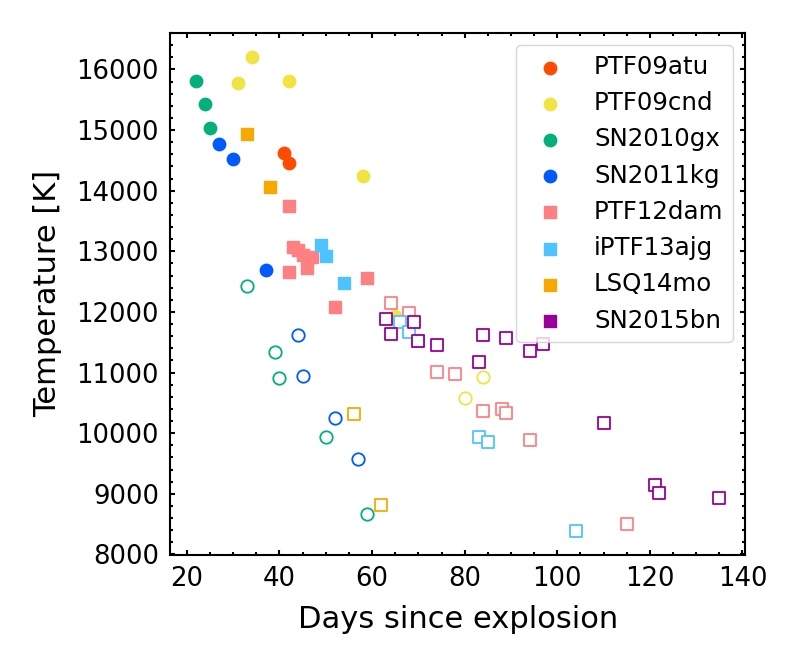}
        \end{center}
    \end{minipage}    
     \begin{minipage}{0.5\hsize}
      \begin{center}
       \includegraphics[width= 0.97\linewidth]{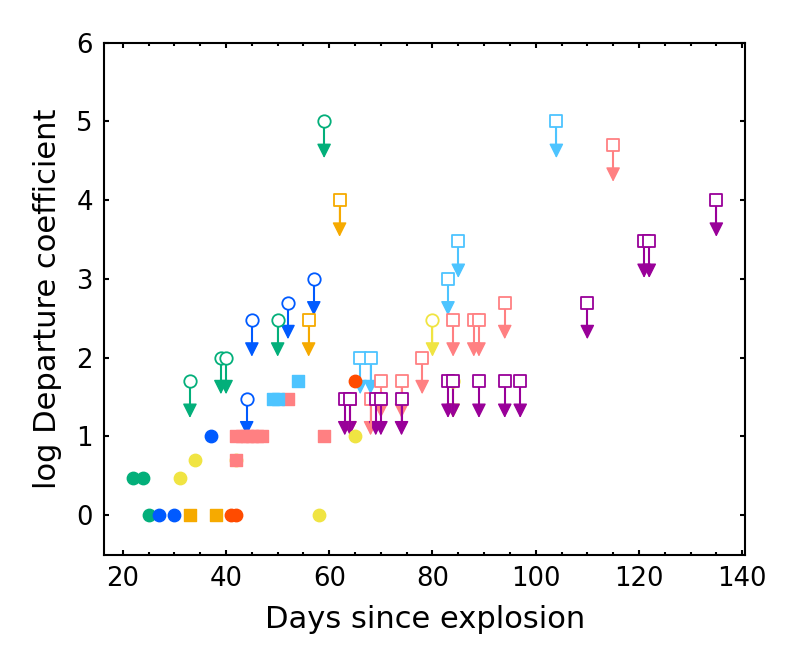}
       \end{center}
    \end{minipage}
  \end{tabular}  
\caption{
Left panel:
Relation between time since explosion and temperature of each spectrum.
The different combinations of the colors and the shapes of the points relate to different SNe as shown in the legend.
The filled and open points represent spectra with and without the W-shaped O II lines, respectively.
Right panel: same as the left panel, but departure coefficient is shown as a function of time since explosion.
Note that the open points show upper limits of the departure coefficients since the respective spectra lack the W-shaped O II lines.
}
  \label{fig:t_vs}
  \end{center}
\end{figure*}

In this section, we show the behavior of the temperature and the departure coefficient as a function of time.
The left panel of Figure \ref{fig:t_vs} shows radiation temperatures just outside the photospheres
in the models with the best parameters as a function of time since explosion.
The temperatures decrease with time.
It is clear that the only spectra with temperatures higher than $T_{\rm R} \sim 12,000$ K show the W-shaped O II lines
(see also \citealt{Konyves-Toth2022}).
Dependence of the W-shaped O II lines on temperature is discussed in more detail in Section \ref{dis:dep}.
The W-shaped O II lines can appear only up to $t_{\rm expl} \sim$ 60 days.
In fact, more luminous objects tend to show the W-shaped O II lines somewhat longer
(see the points with the black edges in Figure \ref{fig:lc}).
This is because, for more luminous objects,
higher temperatures can be achieved for a longer time.

The right panel of Figure \ref{fig:t_vs} shows relations between time since the explosion and the departure coefficients.
The W-shaped O II lines in many of the observed spectra in the early phases ($t_{\rm expl} \ltsim 50$ days)
are reproduced well with the departure coefficients $b_{\rm neb} \sim 1$.
This means that the W-shaped O II lines are formed with little departure from the population in the nebular approximation.
Some of the spectra with the W-shaped O II lines around the phases $t_{\rm expl} \sim 50 - 70$ days
require somewhat larger departure coefficients $b_{\rm neb} \sim 10 - 50$.
For the observed spectra without the W-shaped O II lines, 
we obtained upper limits of the departure coefficients.
When the departure coefficients exceed the upper limits,
the synthetic spectra would produce the W-shaped O II lines.
The upper limits of the departure coefficients become larger at later times.

\begin{figure}[t]
\includegraphics[width= 0.98\linewidth]{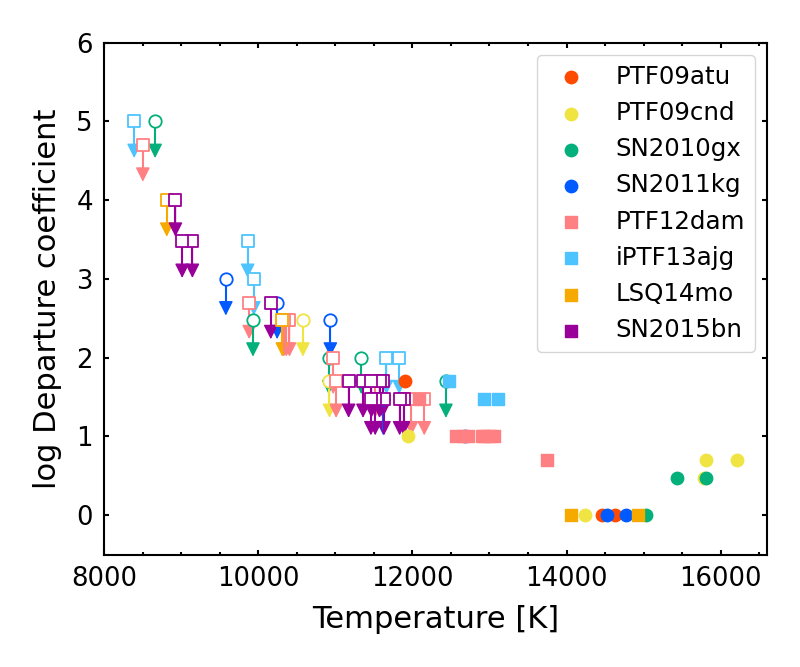}
\caption{Temperature vs departure coefficient. The data and the symbols are the same as in Figure \ref{fig:t_vs}.}
\label{fig:T_b}
\end{figure}

Temperatures and departure coefficients are plotted one against another in Figure \ref{fig:T_b}.
There is a tendency that spectra with higher (lower) temperatures require smaller (larger) departure coefficients.
The spectra with the temperatures $T_{\rm R} \gtsim 14,000$ K show the W-shaped O II lines with the departure coefficient $b_{\rm neb} = 1$ (no departure).
As the temperatures decrease, larger departure coefficients are requisite for the spectra with the W-shaped O II lines.
The departure coefficient for the W-shaped O II lines in the spectra of the SLSNe-I
in our sample is $b_{\rm neb} \sim 50$ at most.
This is in contrast to the departure coefficient for the He I lines in spectra of SNe Ib \citep[$b \sim 10^4 - 10^5$; ][]{Lucy1991}.

\section{Discussion}
\label{sec:dis}

\subsection{Dependence of line strength on temperature}
\label{dis:dep}

\begin{figure}[ht]
\includegraphics[width= \linewidth]{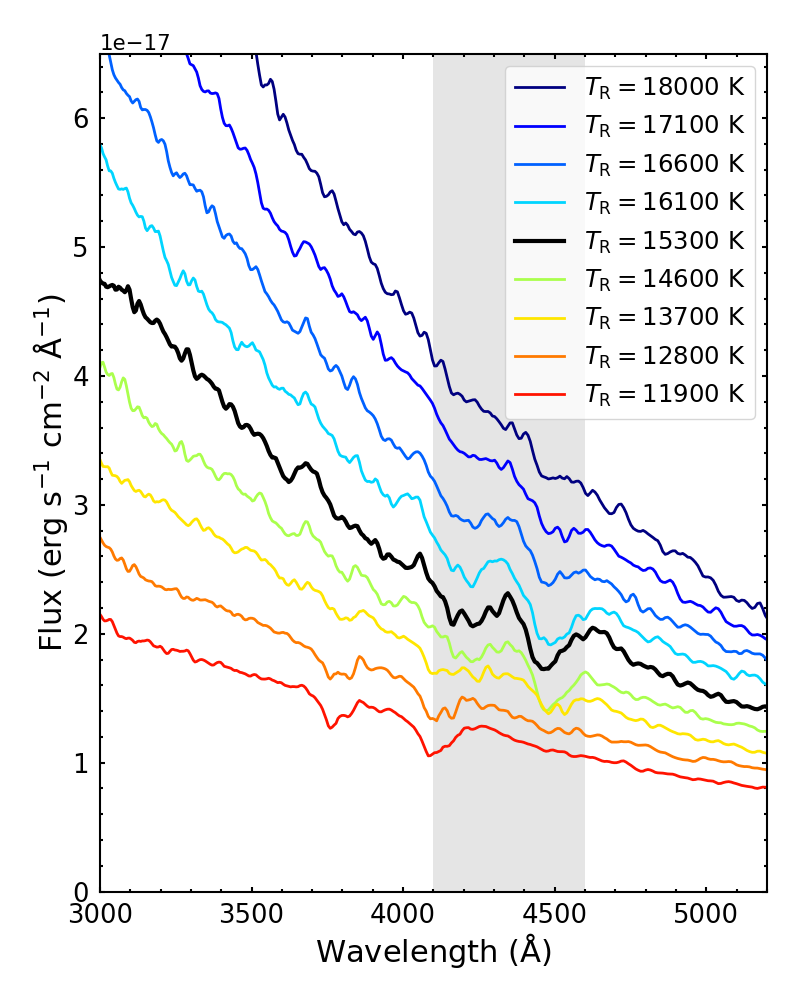}
\caption{
Series of synthetic spectra with various bolometric luminosities $L_{\rm bol}$ for a given combination of the other parameters
to show the gradual influence of temperatures.
Here, the bolometric luminosity $L_{\rm bol}$ is parameterized with a step of log ${L_{\rm bol} = 0.1} \ {\rm erg \ s^{-1}}$
from log ${L_{\rm bol} = 44.2} \ {\rm erg \ s^{-1}}$ (the red line)
to log ${L_{\rm bol} = 45.0} \ {\rm erg \ s^{-1}}$  (the dark blue line).
Other parameters are fixed to $v_{\rm ph}$ = 12,750 ${\rm km \ s^{-1} }$,  $t_{\rm expl}$ = 40 days, and $b_{\rm neb} = 1$.
Accordingly, the density scale factor $f_{\rho}$ is varied from 1.3 to 0.6 so that the dilution factor converges to $1/2$.
The redder and bluer colors show models with lower temperatures (lower bolometric luminosities) and higher temperatures (higher bolometric luminosities).
Radiation temperatures just outside photospheres are shown in the legend.
The black line shows a synthetic spectrum with the deepest W-shaped O II lines ($T_{\rm R} \sim 15,000$ K).
The gray shaded area highlights the W-shaped O II lines.
}
\label{fig:series}
\end{figure}

Here, we discuss the dependence of the strength of the W-shaped O II lines on temperature.
Figure \ref{fig:series} shows a series of synthetic spectra with various
temperatures. For the calculations of the synthetic spectra, the bolometric
luminosity $L_{\rm bol}$ was parameterized to change temperature. The model with
the radiation temperature just outside the photosphere $T_{\rm R} \sim 15,000$ K
shows the strongest W-shaped O II lines. Neither the models with temperatures
$T_{\rm R} \ltsim 13,000$ K nor with $T_{\rm R} \gtsim 18,000$ K show the W-shaped
O II lines.

To understand this behavior, we evaluate the strength of the absorption lines from
the Sobolev optical depth as in Equation (\ref{eq:tau}) under the
nebular approximation (see \citet{Hatano1999} for a similar analysis for normal SNe under LTE).
Absorption lines can appear when the Sobolev optical depth is $\tau_{lu}$ \gtsim 1 
above the photosphere over a range of velocities.
A typical velocity range of the line forming region is $v \sim v_{\rm ph} - 1.5 \ v_{\rm ph}$
to reproduce the width of the absorption line seen in the O II lines.

\begin{figure*}[th]
\begin{center}
  \begin{tabular}{c}
    \begin{minipage}{0.5\hsize}
      \begin{center}
       \includegraphics[width= 0.97\linewidth]{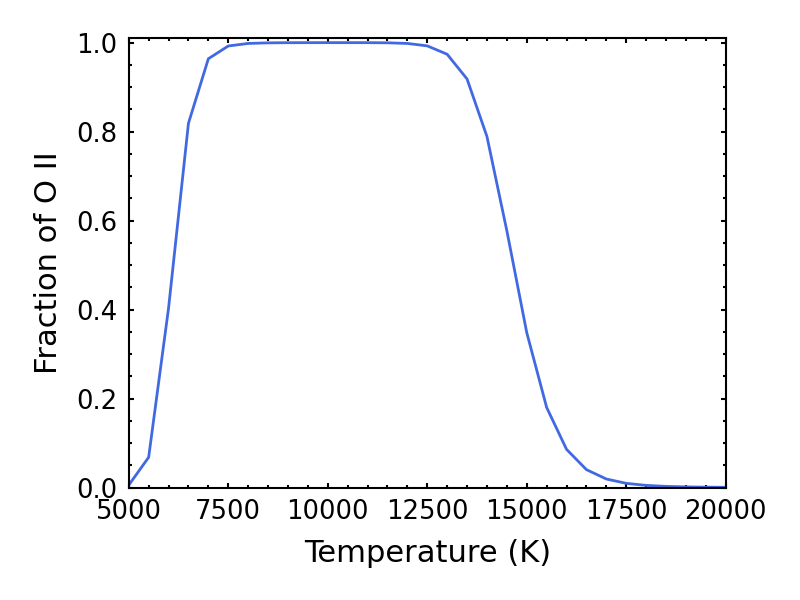}
        \end{center}
    \end{minipage}    
     \begin{minipage}{0.5\hsize}
      \begin{center}
       \includegraphics[width= 0.97\linewidth]{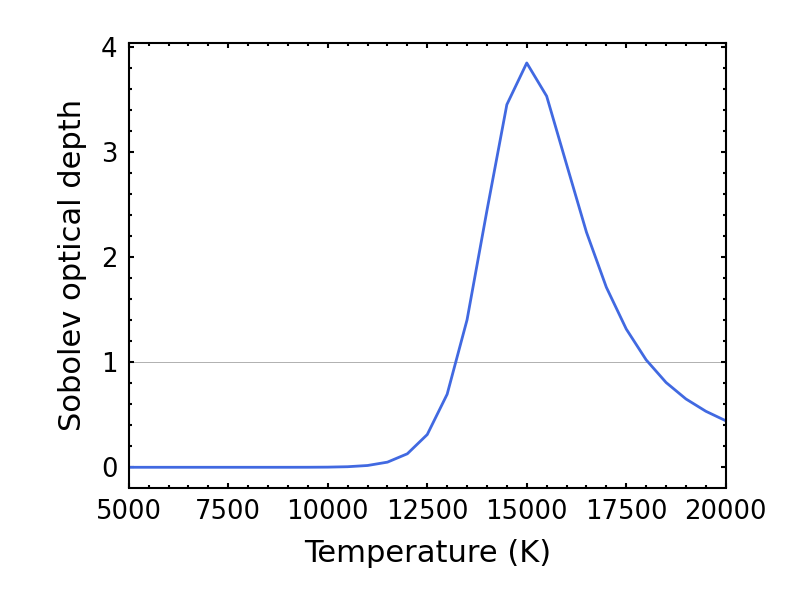}
       \end{center}
    \end{minipage}
  \end{tabular}  
\caption{
Fraction of O II among and the Sobolev optical depth at typical photospheres ($W=1/2$ and $\rho = 3 \times 10^{-14} {\rm \ g \ cm^{-3}}$)
with the same abundance as that adopted to the calculations of the synthetic spectra in Section \ref{sec:setup}.
Left panel: dependence of the fraction of O II on temperature, estimated by Equation (\ref{eq:ionize}).
Right panel: dependence of the Sobolev optical depth of one of the most prominent lines among the W-shaped O II lines
estimated by Equation (\ref{eq:excite}) and (\ref{eq:tau}),
where 
$f_{lu} = 0.34$ \citep{Wiese1996},
$t_{\rm expl} = 40$ days,
and $\lambda_{lu} = 4,649 \ {\rm \AA}$.
The horizontal gray line in the right panel marks the Sobolev optical depth $\tau_{lu} = 1$, above which the W-shaped O II lines can appear.
}
  \label{fig:tau}
  \end{center}
\end{figure*}

The left panel of Figure \ref{fig:tau} shows the fraction of O II at typical
photospheric parameters evaluated with the same abundance adopted for the spectral
synthesis calculations in Section \ref{sec:setup}. Near the photospheres, density
is typically $\rho = 3 \times 10^{-14} {\rm \ g \ cm^{-3}}$, which is almost
constant over time. The right panel of Figure \ref{fig:tau} shows the Sobolev
optical depth $\tau_{lu}$ of the W-shaped O II lines at the photosphere obtained
from the ionization fraction and Equation (\ref{eq:excite}) and (\ref{eq:tau}). The
Sobolev optical depth shown here is that of one of the most prominent O II lines, 
$\lambda_{lu} = 4,649 \ {\rm \AA}$. For the line, we apply $f_{lu} =
0.34$ \citep{Wiese1996} and $t_{\rm expl} = 40$ days in Equation (\ref{eq:tau}).

The Sobolev optical depth peaks at $T_{\rm R} \sim 15,000$ K with a value
$\tau_{lu} \sim 4$. As the temperature decreases from $T_{\rm R} \sim 15,000$ K,
the Sobolev optical depth steeply decreases since the number density of the excited
state decreases with the Boltzmann factor. On the other hand,  as the temperature
increases from $T_{\rm R} \sim 15,000$ K, the Sobolev optical depth decreases
because O II is ionized to O III. Therefore, a temperature around $T_{\rm R} \sim
15,000$ K makes the Sobolev optical depth peak. Only at $T_{\rm R} \sim 13,000 -
18,000$ K ($T_{\rm R} \sim14,000 - 16,000$ K), the Sobolev optical depth exceeds
$\tau_{lu} \sim 1$ ($\tau_{lu} \sim 3$).

This dependence of line strength on temperature can also be seen in Figure
\ref{fig:T_b}. The synthetic spectra with higher temperatures than $T_{\rm R} \sim
14,000$ K show the W-shaped O II lines  with a departure coefficient $b_{\rm neb}
= 1$. On the other hand, the synthetic spectra with temperatures lower than $T_{\rm
R} \sim 14,000$ K  require departure ($b_{\rm neb} > 1$) to show the W-shaped O II
lines. Most of the observed spectra with temperatures $T_{\rm R} \ltsim 12,000$ K
do not show the W-shaped O II lines.

Although SLSNe-I are sometimes classified into two types based on the absence or presence of the W-shaped O II lines in their spectra
\citep[e.g.,][]{Konyves-Toth2021},
our results indicate that the difference between the two types would be only temperature.
It is natural that all the spectra of SN 2015bn do not show the W-shaped O II lines because of its low temperatures.

\subsection{Effects of host extinction on departure coefficients}
\label{dis:13ajg}

We now examine the effect of host extinction on the departure coefficients via temperature.
Although the temperatures estimated in the model are affected by host extinction,
it was not corrected for as mentioned in Section \ref{sec:obs}.
Here, we investigate a possible host extinction of iPTF13ajg,
which requires one of the largest departure coefficients ($b_{\rm neb} \ltsim 50$) among our sample.
The upper limit of the host extinction of iPTF13ajg is estimated to be $E(B - V) < 0.12$
from the absence of Na I D lines \citep{Vreeswijk2014}.
Thus, we applied $E(B - V) = 0.05$ to the second spectrum of iPTF13ajg taken at MJD 56391 (one of the best spectra).

\begin{figure}[ht]
\includegraphics[width= 1.0\linewidth]{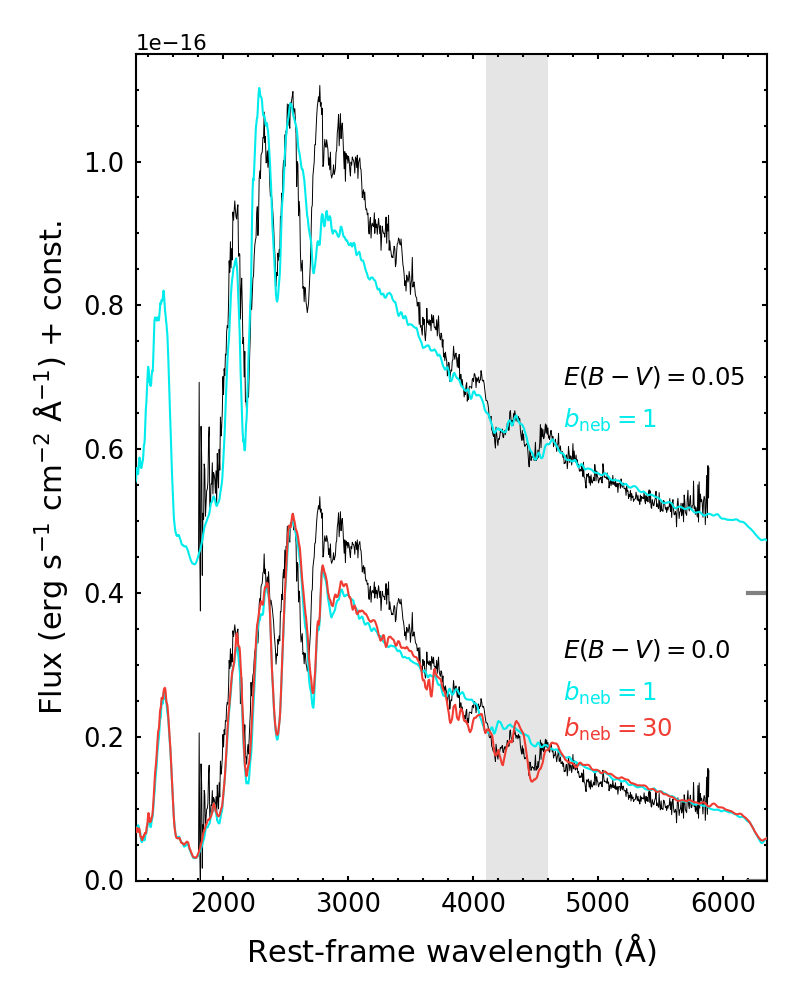}
\caption{
Modeling of observed spectra of iPTF13ajg taken at MJD 56391 with
the host extinction $E(B - V) = 0.0$ (the lower spectrum)
and $E(B - V) = 0.05$ (the upper spectrum) corrected.
The black lines show the observed spectra.
The red line shows the model for the spectrum with the host extinction $E(B - V) = 0.0$ with the best parameters:
$f_\rho = 1.0$,
$L_{\rm bol} = 10^{44.5} {\rm \ erg \ s^{-1}}$,
$v_{\rm ph} = 12,250 {\rm \ km \ s^{-1}}$,
$t_{\rm expl} = 50 $ days, and
$b_{\rm neb} = 30$.
The model with $b_{\rm neb}=1$ and the other parameters same as above is shown with the lower light blue line.
The upper light blue line shows the model for the spectrum with the host extinction $E(B - V) = 0.05$ with the best parameters:
$L_{\rm bol} = 10^{44.6} {\rm \ erg \ s^{-1}}$
$t_{\rm expl} = 47 $ days,
$b_{\rm neb} = 1$, and 
the other parameters are the same as above.
The gray shaded area highlights the W-shaped O II lines.
Vertical offsets for each spectrum are shown by the horizontal lines on the right y-axis.
}
\label{fig:13ajg}
\end{figure}

The host-extinction-corrected spectra and the models with the best parameters are shown in Figure \ref{fig:13ajg}.
For the model of the spectrum with $E(B - V) = 0.0$ (before the host extinction correction),
the best parameters are
$f_\rho = 1.0$,
$L_{\rm bol} = 10^{44.5} {\rm \ erg \ s^{-1}}$,
$v_{\rm ph} = 12,250 {\rm \ km \ s^{-1}}$,
$t_{\rm expl} = 50 $ days,
and $b_{\rm neb} = 30$.
These parameters give a radiation temperature just outside the photosphere $T_{\rm R} \sim 13,000$ K.When we assume $E(B - V) = 0.05$,
the intrinsic spectrum becomes brighter and bluer.
Then, the best parameters are
$f_\rho = 1.0$,
$L_{\rm bol} = 10^{44.6} {\rm \ erg \ s^{-1}}$,
$v_{\rm ph} = 12,250 {\rm \ km \ s^{-1}}$,
$t_{\rm expl} = 47 $ days,
and $b_{\rm neb} = 1$.
These parameters give a radiation temperature just outside the photosphere $T_{\rm R} \sim 14,500$ K.
By this somewhat higher temperature, the model spectrum with $b_{\rm neb} = 1$ produces
a deeper W-shaped O II absorption feature.

These results demonstrate that the required departure coefficients is quite sensitive to the host extinction, which is always difficult to estimate.
Even a small host extinction correction largely affects the UV fluxes, increasing the estimated temperature.
Thus, the required departure coefficient tends to be smaller when the host extinction is applied.

\subsection{Constraints on ionization rate and implication to power sources} 
\label{dis:power}

We here examine non-thermal processes in ejecta of SLSNe
despite our finding that population of the excited states of O II does not largely deviate from population in the nebular approximation. 
This may yield constraints on the power source of SLSNe-I
because population would be influenced by non-thermal excitation/ionization to some extent
whenever there are $\gamma$-rays from either \Nifs \ or a magnetar in ejecta.
We test this for temperatures of $T = 8,000 - 12,000$ K,
where the nebular approximation does not lead to formation of the W-shaped O II lines.

In analogy to He I lines in spectra of SNe Ib,
the ionization rate can be constrained by the population of the excited states of O II.
For the appearance of the He I lines,
the excited states of He I are populated by high energy electrons
that are produced by Compton scattering of $\gamma$-rays from \Nifs \ decay \citep{Lucy1991, Hachinger2012}.
The high energy electrons ionize He I to He II with a rate determined by the energy deposition rate of $\gamma$-rays from \Nifs \ decay.
Then, He II recombines to the excited states of He I \citep[see Figure 1 of][]{Tarumi2023}.
Therefore, the population of the excited states of He I is related to the non-thermal ionization rate.
While the presence of He I lines in spectra of SNe Ib gives the non-thermal ionization rate,
the absence of the W-shaped O II lines in spectra of SLSNe-I can give
an upper limit to the non-thermal ionization rate through population of the excited states of O II.

The number density of the excited states of O II should not exceed a certain value that produces the W-shaped O II lines.
The typical value of the Sobolev optical depth producing absorption lines 
is of the order of $\tau_{lu} \sim 1$.
This corresponds to densities $n_{i, j, l} \gtsim 1 {\rm \ cm^{-3}}$ for the W-shaped O II lines
obtained from Equation \ref{eq:tau} by adopting
$\lambda_{lu} = 4,649 \ {\rm \AA}$,
$f_{lu} = 0.34$ \citep{Wiese1996},
and $t_{\rm expl} = 40$ days.
This number density is discussed further below and shown to give an upper limit of the ionization rate.
The ionization rate can then be converted to an energy deposition rate
through the fraction of the deposition energy spent for ionization
(among ionization, excitation, and heating of thermal electrons, see e.g., \citet{Fransson1989}),
which is called work per ion pair.
Finally, this energy deposition rate can be related to power sources
through the $\gamma$-ray spectra of power sources (and $\gamma$-ray transport).

\begin{figure}[th]
\centering
\includegraphics[width= 0.98\linewidth,  bb= -10 -5 780 1020]{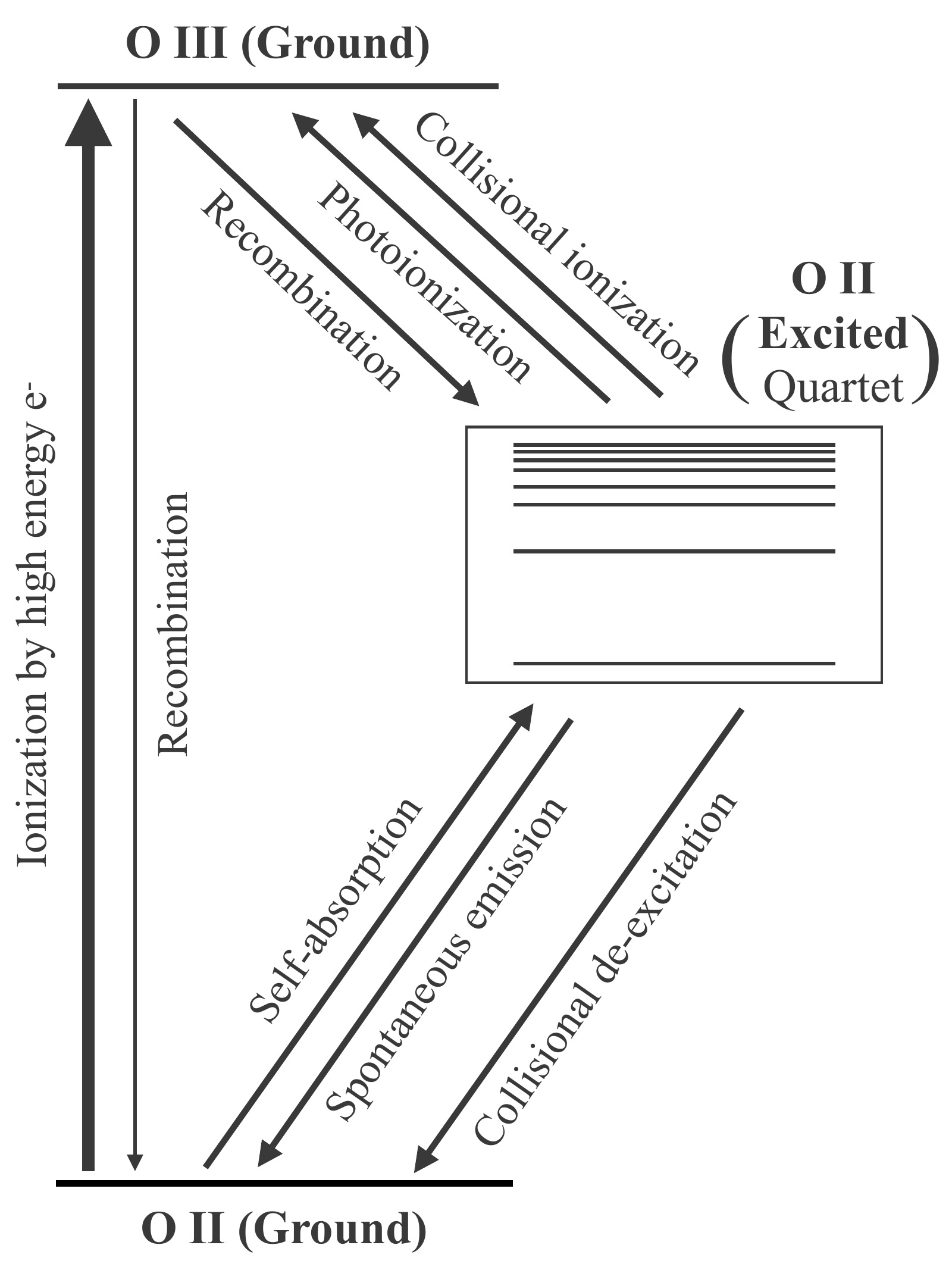}
\caption{
A schematic diagram showing the populating mechanism of the excited O II spin-quartet state.
Here, we consider 3s $^4$P,
which absorbs photons with 4,349 $\rm \AA$ and 4,649 $\rm \AA$ (see also Figure \ref{fig:config}).
}
\label{fig:rate}
\end{figure}

In order to estimate the population of the excited states of O II under certain ionization and excitation conditions,
we solve a simplified rate equation as illustrated in Figure \ref{fig:rate}.
We only consider three levels:
the ground state of O II, 
the ground state of O III (35.1 eV higher than the ground state of O II),
and the excited O II spin-quartet state (23.0 eV higher than the ground state of O II).
We do not consider the case of the excited state of O II in the spin-doublet state,
which is expected to be similar to that in the spin-quartet state.
This is because both the excited states would be balanced to be comparably occupied via O III.
This is also seen in the excited state of He I in the spin-singlet state and that in the spin-triplet state
achieved by the balance via He II \citep{Lucy1991}.
Also, we separately solve two equations below:
an equation for the balance between the ground state of O II and the ground state of O III,
and an equation for inflow to and outflow from the excited O II spin-quartet state.

We first consider balance between the ground state of O II and that of O III by non-thermal processes.
The balance between the ionization from O II to O III and the recombination from O III to O II gives the number density of O III $n_{\rm OIII}$:
\begin{equation}
\label{eq:rate1}
\beta n_{\rm OII} = n_{\rm OIII} n_{\rm e} \alpha_{\rm O III \rightarrow O II} (T),
\end{equation}
where
$\beta$ is the non-thermal ionization rate $({\rm s^{-1}})$,
$n_{\rm OII}$ is the number density of the ground state of O II,
$n_{\rm e}$ is the number density of electrons,
and 
$\alpha_{\rm O III \rightarrow O II} (T)$ is an effective recombination rate from the ground state of O III to the ground state of O II.
Since we consider only non-thermal ionization here,
the ionization rate $\beta$ is expressed as
\begin{equation}
\label{eq:beta}
\beta = \frac{\dot{e}}{w},
\end{equation}
where
$\dot{e}$ is the energy deposition rate per particle given by power sources $({\rm erg \ s^{-1}})$,
and $w$ is the energy (erg) required to ionize an ion (work per ion pair).
Here, $w$ is usually obtained from the Spencer-Fano equation \citep{Spencer1954, Kozma1992}.

Treatment of the recombination processes is not trivial.
The direct recombination to the ground state is often largely suppressed 
because it is immediately followed by absorption.
In fact, in a typical density we consider, the optical depth of
the recombination photons is quite high ($\tau > 10^6$).
In a realistic ejecta including various elements other than O,
the recombination photons can also be absorbed by other ions,
and thus, some direct recombination can still occur.
Another path is recombination through the excited states.
However, since we do not solve a full rate equation including
the excited states of O II (see below),
it is difficult to accurately evaluate the effective recombination rate
through many excited states.
Under these circumstances, we approximately adopt the direct recombination rate
to the ground state from \citet{Nahar1999} as an effective recombination rate
 $\alpha_{\rm O III \rightarrow O II} (T)$ in Equation (\ref{eq:rate1}).
This corresponds to the limit of the most efficient recombination to O II.
In reality, the recombination rate would be lower than what is adopted here,
and the number density of O III ions would be enhanced (implications of this assumption are discussed below).

Next, we solve the equation for inflow to and outflow from the excited O II spin-quartet state.
This gives the number density of O II in that state $n_{\rm OII, ex}$:
\begin{align}
\label{eq:rate2}
 \begin{split}
n_{\rm OIII}  n_{\rm e} \alpha_{\rm OIII \rightarrow OII, ex} & (T) \\
=
n_{\rm OII, ex} 
(R_{\rm PI} +
& n_{\rm e}q_{\rm OII, ex \rightarrow OII} + \\
&
n_{\rm e}q_{\rm OII, ex \rightarrow OIII} +
\beta_{\rm esc} A_{\rm ex \rightarrow gs}),
 \end{split}
\end{align}
where
$\alpha_{\rm OIII \rightarrow OII, ex}(T)$ is the recombination rate from the ground state of O III to the excited O II spin-quartet state,
$R_{\rm PI}$ is a photoionization rate,
$q_{\rm OII, ex \rightarrow OII}$  is the collisional transition rate from the excited O II spin-quartet state to the ground state of O II,
$q_{\rm OII, ex \rightarrow OIII}$ is the collisional transition rate from the excited O II spin-quartet state to the ground state of O III,
$\beta_{\rm esc}$ is the Sobolev escape probability,
and 
$A_{\rm ex \rightarrow gs}$ is the Einstein coefficient for the transition from the excited O II spin-quartet state to the ground state of O II.
Note that the last term on the right side of Equation \ref{eq:rate2} includes both 
self-absorption and spontaneous emission.
The photoionization rate $R_{\rm PI}$ is computed as 
\begin{equation}
\label{eq:R_PI}
R_{\rm PI}
=
\int_{h \nu = {\rm 12.1 \ eV}}^{\infty}
\sigma_{\rm PI}(\nu)
\frac
{W B_\nu (T)}
{h\nu}
d\nu,
\end{equation}
where $\sigma_{\rm PI} (\nu)$ is a photoionization cross section.
The transition rate by electron collisions $q_{if}$ is computed as
\begin{eqnarray}
\label{eq:q_if}
q_{i \rightarrow f} =
 \begin{dcases}
      \frac{8.63 \times 10^{-6}}{g_i \sqrt{T}} \gamma_{if} e^{-(E_f- E_i)/kT}  & (E_i < E_f)\\
      \frac{8.63 \times 10^{-6}}{g_i \sqrt{T}} \gamma_{if}                            & (E_i > E_f)
 \end{dcases}
\end{eqnarray}
\citep{Eissner1969},
where
$g_i$ is the statistical weight of an initial state,
$\gamma_{if}$ is an effective collision strength from an initial state to a final state,
and 
$E_i$ and $E_f$ are an energy level of an initial state and a final state, respectively.
The escape probability $\beta_{\rm esc}$ is computed as 
\begin{equation}
\label{eq:esc}
\beta_{\rm esc} = \frac{1 - e^{-\tau_{lu}} }{\tau_{lu}}
\end{equation}
\citep{Castor1970}
with the Sobolev optical depth $\tau_{lu}$ in Equation (\ref{eq:tau}).

\begin{figure}[t]
\centering
\includegraphics[width= 0.98\linewidth]{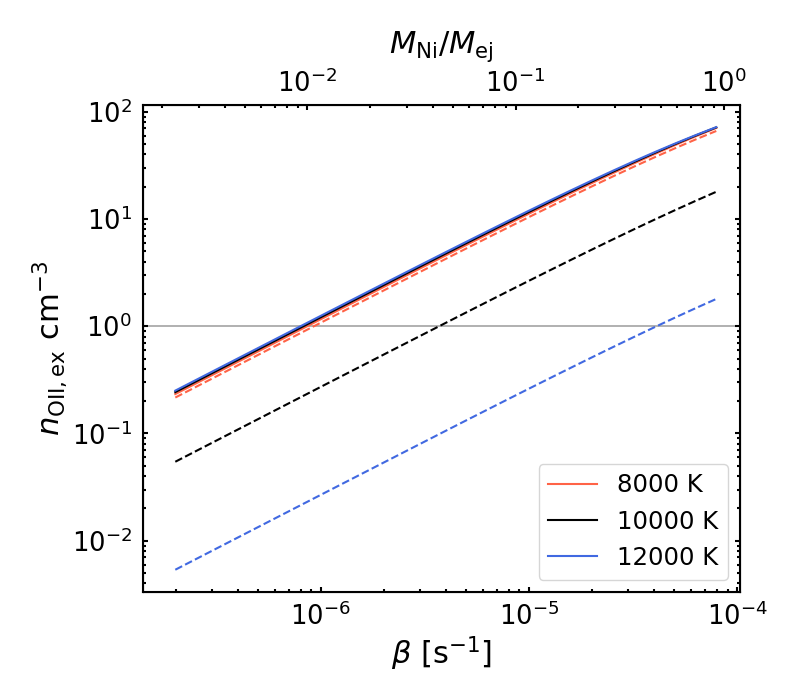}
\caption{
The number densities for the excited O II spin-quartet state $n_{\rm OII, ex}$ as a function of the ionization rate (the lower x-axis).
The upper x-axis shows the $M_{\rm Ni} / M_{\rm ej}$ ratio corresponding to the ionization rate.
Different colors show different temperatures, as shown in the legend.
Within this range of temperatures,
O II is secured to be dominant as shown in the left panel of Figure \ref{fig:tau}
with typical density $\rho \sim 3 \times 10^{-14} {\ \rm g \ cm^{-3}}$ at a photosphere.
The number densities for the excited O II spin-quartet state, $n_{\rm OII, ex}$, in the nebular approximation at the temperature shown
are too small to produce the W-shaped O II lines.
The dashed lines and solid lines show the number densities with and without photoionization from the excited O II spin-quartet state.
All solid lines (the cases without photoionization) overlap near the dashed line for $T = 8,000$ K.
The gray line represents $n_{\rm OII, ex} = 1 {\ \rm cm^{-3}}$, beyond which the W-shaped OII lines can appear.
}
\label{fig:n_OII_ex}
\end{figure}

The procedure above gives the number density of the excited stated of O II $n_{\rm OII, ex}$
as a function of the ionization rate (Figure \ref{fig:n_OII_ex}).
Here, we assumed that all the ejecta consist of O, and that O II is dominant as shown in the left panel of Figure \ref{fig:tau}.
This provides a condition of $n_{\rm e} = n_{\rm O II} + 2 n_{\rm O III}$.
Also, we adopted the following parameters:
$\rho = 3 \times 10^{-14} {\ \rm{g} \ cm^{-3}}$ as a typical density near the photosphere,
$\alpha_{\rm OIII \rightarrow OII, ex}(T)$ from \citet{Nahar1999},
$\sigma_{\rm PI} = 10^{-17} {\ \rm cm{^2}}$ \citep{Nahar1999},
$g_i = 4$,
$\gamma_{if} =1$,
$E_i = 23.0$ eV,
$E_f = 0.0$ eV for the collisional de-excitation, $E_f = 35.1$ eV for the collisional ionization, 
$A_{\rm ex \rightarrow gs} = 9.8 \times 10^8 {\ \rm s^{-1}}$ \citep{Nahar2010},
and $W = 1/2$;
for $\tau_{lu}$,
$f_{lu} = 4.3 \times 10^{-2}$ \citep{Wiese1996},
$n_{i,j,l} = n_{\rm O II}$,
$\lambda_{lu} = 539 {\ \rm \AA}$,
and $t_{\rm expl} = 40$ days.
Note that the photoionization cross section is assumed to be constant 
because the ionization cross section has many resonances just above the ionization energy.
As the peak of the blackbody radiation with the temperature $T = 8,000 -  12,000$ K is located at much lower energy than the ionization threshold,
the assumption of the constant cross section is effectively applied just above the ionization threshold.

Figure \ref{fig:n_OII_ex} shows the number densities for the excited O II spin-quartet near the photosphere as a function of the ionization rate.
The number density of the excited O II spin-quartet exceeds the critical density $n_{\rm OII, ex} \sim 1 {\ \rm cm^{-3}}$
(the Sobolev optical depth $\tau_{lu} \sim 1$)
when the ionization rate is $\beta \sim 3 \times 10^{-6} {\ \rm s^{-1}}$ at $T = 10,000$ K.
At $T = 8,000$ K and $T = 12,000$ K,
the number density for the excited state of O II exceeds the critical density
when the ionization rate is $\beta \sim 10^{-6} {\ \rm s^{-1}} {\ \rm and \ } \beta \sim 3 \times 10^{-5} {\ \rm s^{-1}}$, respectively.

We note that the estimated number densities for the excited state of O II are lower-limit estimates
because we assume blackbody radiation even at short wavelengths.
In actual SNe, photons at short wavelengths are suppressed because of line blanketing.
Smaller photoionization rates should increase the estimated number densities of the excited stated of O II.
To show this effect, we compare the number densities for the excited O II spin-quartet with photoionization rate to those without photoionization
since the photoionization rate is uncertain because of line blanketing.
The results at all the temperatures $T = 8,000 {\ \rm K}, \ 10,000 {\rm \ K, and \ } 12,000 {\ \rm K}$ without photoionization are almost the same
as the result at $T = 8,000 {\ \rm K}$ with photoionization.
At the temperatures $T = 10,000$ K and $12,000$ K, 
the effect of the photoionization is the strongest to de-populate the excited state of O II in comparison to all the other processes
(collisional de-excitation, collisional ionization, and spontaneous emission with self-absorption). 
In light of line blanketing, the result with radiation temperature $T = 8,000$ K may best represent the actual condition in the ejecta of SLSNe.

The ionization rate can be translated to the energy deposition rate if the work per ion pair in Equation (\ref{eq:beta}) is given.
The work per ion pair becomes larger when the electron fraction is larger,
since a larger fraction of the deposition energy is spent for heating of thermal electrons, not for ionization \citep{Kozma1992}.
In electron-rich environments where ionization of metals produces many electrons,
a typical value of the work per ion pair (calculated for Type Ia SNe) is $w \sim 30 \ I$, where $I$ is an ionization potential for an ion \citep[]{Axelrod1980}.
Hereafter, we assume a work per ion pair of $w = 30 \ I$ for ionization of O II to O III in SLSNe-I.
The deposition rates then corresponding to an ionization rate $\beta$ that makes  $n_{\rm OII, ex} \gtsim 1 {\ \rm cm^{-3}}$
are
$\dot{e} \sim 2 \times 10^{-15} {\ \rm erg \ s^{-1}}$,
$\dot{e} \sim 5 \times 10^{-15} {\ \rm erg \ s^{-1}}$, and
$\dot{e} \sim 5 \times 10^{-14} {\ \rm erg \ s^{-1}}$
at $T = 8,000$ K, $T = 10,000$ K, and $T = 12,000$ K,
respectively.

Finally, we demonstrate that the energy deposition rate can be related to power sources
if the $\gamma$-ray spectrum emitted by the power sources and subsequent $\gamma$-ray transport are known.
Here, to give crude constraints on the mass of \Nifs \ in SLSNe-I,
we hypothetically assume
that \Nifs \ is the power source for SLSNe-I.
For simplicity,
\Nifs \ is assumed to uniformly supply the deposition energy to the ejecta (full mixing).
The optical depth of $\gamma$-rays is $\tau_{\gamma} \gtsim 1$ in photospheric phases.
Thus, for fully mixed \Nifs \ distribution,
the energy deposition rate given by \Nifs \ decay can be approximately written as
\begin{eqnarray}
\label{eq:e_dot}
\dot{e} \simeq \frac{L_{\rm decay}}{N},
\end{eqnarray}
where $L_{\rm decay}$ is the decay luminosity of \Nifs \ ${(\rm erg \ s^{-1}})$
and $N$ is the number of atoms in the ejecta.
The decay luminosity is given as \citep[e.g.,][]{Nadyozhin1994}:
\begin{align}
\label{eq:L_decay}
\begin{split}
L_{\rm decay}  (t_{\rm expl}) 
=
(
6.45 \ & e^{- \frac{t_{\rm expl}}{8.8 {\ \rm d}}}
 +  1.45 \ e^{- \frac{t_{\rm expl}}{111.3 {\ \rm d}}}
) 
\\
&
\times10^{43} {\ \rm{erg \ s^{-1}}}
\left(
\frac {M_{\rm Ni}}{M_\odot}
\right),
\end{split}
\end{align}
where $M_{\rm Ni}$ is the mass of \Nifs.
Here, we apply $t_{\rm expl} = 40$ days.
The number of atoms is $N_{\rm A} = M_{\rm ej}/ \langle A \rangle m_{\rm p}$,
where
$M_{\rm ej}$ is an ejecta mass,
$\langle A \rangle$ is the average mass number in the ejecta (here, $\langle A \rangle = 16$ for pure O),
and $m_{\rm p}$ is the mass of a proton.
The ejecta mass is fixed to a typical value for SLSNe-I, $M_{\rm ej} \sim 6 \ M_\odot$ \citep{Blanchard2020}.
Then, the deposition rate needed for appropriate ionization rates
can be converted to a mass ratio of \Nifs \ vs the ejecta $M_{\rm Ni}/M_{\rm ej}$.
This ratio is shown as to the upper x-axis in Figure \ref{fig:n_OII_ex}.
The mass ratio corresponding to the ionization rate that makes  $n_{\rm OII, ex} \gtsim 1 {\ \rm cm^{-3}}$
is
$M_{\rm Ni}/M_{\rm ej} \sim 0.05$ at  $T = 10,000$ K.
This is translated to \Nifs \ mass of $M_{\rm Ni} \sim 0.3 \ M_\odot$ for $M_{\rm ej} = 6 \ M_\odot$.
If SLSNe-I are entirely powered by \Nifs, the required \Nifs \ mass is $\gtsim 3 \ M_\odot$,
but it would lead to too strong a non-thermal ionization.

We emphasize that our estimates above involve a number of assumptions and uncertainties,
and the upper limit of the \Nifs \ mass is still quite uncertain.
For example, the recombination rate adopted in the ionization balance
(i.e., the direct recombination rate to the ground state) is expected to be
lower because of immediate reabsorption.
A lower recombination rate would increase the number density of O III ions
and enhance the population of the excited state of O II,
giving stronger W-shaped O II lines.
Then, the upper limit of \Nifs\ would become smaller (or tighter).
On the other hand, there may also be the effects making the upper
limit of the \Nifs \ mass higher.
For example, we only evaluate the Sobolev optical depth just outside the photosphere.
But, in reality, the line is formed over a velocity range of up to $v \sim 1.5 \ v_{\rm ph}$
(Section \ref{dis:dep}).
To have $\tau_{lu} > 1$ for a wider velocity region,
more \Nifs \ would probably be required, making the upper limit of the \Nifs\ mass higher.
Also, we assumed full mixing to translate ionization rate to the \Nifs\ mass.
Under less mixing, energy of non-thermal electrons is spent inside the photosphere,
which decreases the non-thermal ionization rate outside the photosphere.
Thus, with less mixing, the upper limit of the \Nifs\ mass would also become higher (i.e., less stringent).

There are also other assumptions/uncertainties:
the simplified rate equation,
the uncertain photoionization rate because of the line blanketing,
and the uncertain work per ion pair in ejecta of SLSNe-I.
Nevertheless, our work demonstrates that spectroscopic properties can be in principle used to give constraints on the mass of \Nifs \ in SLSNe-I,
which roughly corresponds to the mass ratio $M_{\rm Ni}/M_{\rm ej}$ of the order of 0.1.
In addition, in the cases where SLSNe-I are powered by magnetars,
spectroscopic properties can also be used to give constraints on spectra of $\gamma$-rays from magnetars via ionization rates
if combined with detailed transfer calculations \citep{Vurm2021, Murase2021}.

\section{Summary}
\label{sec:sum}

We have performed systematic spectral calculations to model the observed spectra of eight SLSNe-I
to quantify the conditions for the formation of the W-shaped O II lines.
We find that many of the pre-/near-maximum spectra with the W-shaped O II lines
can be reproduced well with the departure coefficient $b_{\rm neb} \sim 1$ (i.e., without departure)
at the temperatures $T_{\rm R} \sim 14,000 - 16,000$ K near the photosphere.
This suggests that departure from nebular-approximation conditions
is not necessarily large for the formation of the W-shaped O II lines in spectra of SLSNe-I.
The appearance of the W-shaped O II lines is very sensitive to temperature.
Thus, to understand the physical conditions for the line formation,
it is important to estimate accurately the temperature of the ejecta
from spectral modeling (rather than a simple fitting of the spectral energy distribution).
We also highlight the importance of the extinction correction in the host galaxy;
even a small extinction correction ($E(B-V) \sim 0.05$) can increase the intrinsic UV fluxes,
which tends to increase the estimated temperatures by $\sim 2,000 - 3,000$ K.

Finally, we have shown that the absence of the the W-shaped O II lines
in spectra with a lower temperature ($T \ltsim 12,000$ K) can be exploited to constrain the non-thermal ionization rate in the ejecta.
Solving the simplified rate equation gives an upper limit to the non-thermal ionization rates.
Under the several assumptions,
this upper limit is roughly translated to an upper limit on the mass ratio $M_{\rm Ni}/M_{\rm ej}$ of order 0.1.
Similar methods can also be applied to give constraints on $\gamma$-ray spectra
of magnetars if detailed $\gamma$-ray transport is considered.
Although our estimate involves a number of assumptions for simplification,
our work demonstrates that spectroscopic properties can be used
to give independent constraints on the power sources of SLSNe-I.

\acknowledgements
We would like to thank K. Kashiyama, S. Kimura, T. Nagao, and C. Fransson for fruitful discussions.
S.S. is supported by JSPS (Japan Society for the Promotion of Science) Research Fellowship for Young Scientists (21J22515) and GP-PU (Graduate Program on Physics for the Universe), Tohoku University.
This research is partly supported by JST FOREST Program (grant No. JPMJFR212Y, JPMJFR2136) and JSPS KAKENHI grant JP21H04997, JP23H00127, JP23H04894, and 23H05432.


\appendix

\renewcommand{\thetable}{A}

\section{Results of spectral modeling}
\label{app:param}
The best parameters of the models for all the available observed spectra are summarized in Table \ref{table:param}.

\startlongtable
\begin{deluxetable*}{ccc|ccccc|cc}

\tablecaption{Summary of the best-fit parameters for each spectrum.} 
\tablehead{Name & MJD & W-shaped O II lines & $f_{\rm rho}$ & log $L_{\rm bol} ~ ({\rm erg ~ s^{-1}})$  & $v_{\rm ph} \ ({\rm km ~ s^{-1}})$  & $t_{\rm expl}$ (days) &   $b_{\rm neb}$
    & $T_{\rm R}$ (K)}
\startdata
PTF09atu & 55032 & Yes & 0.5 & 44.35 & 11000 & 41 &  1 & 14600 \\ 
                & 55034 & Yes &      & 44.35 & 10750 & 42 &  1 & 14500 \\ 
                & 55068 & Yes &      & 44.40 & 9750 & 65 &  50 & 11900 \\ 
PTF09cnd & 55055 & Yes & 0.7 & 44.50 & 13500 & 31 &  3 & 15800 \\ 
                & 55059 & Yes &       & 44.60 & 13000 & 34 &  5 & 16200 \\ 
               & 55068 & Yes &       & 44.65 & 12000 & 42 &  5 & 15800 \\ 
               & 55089 & Yes &      & 44.65 & 10750 & 58 &  1 & 14200 \\ 
               & 55097 & Yes &      & 44.45 & 10250 & 65 &  10 & 11900 \\ 
               & 55116 & No &      & 44.30 & 9000 & 80 &  $<$ 300 & 10600 \\ 
               & 55121 & No &      & 44.35 & 8750 & 84 &  $<$ 50 & 10900 \\ 
SN2010gx & 55273 & Yes & 3.0 & 44.50 & 18500 & 22 &  3 & 15800 \\ 
                 & 55276 & Yes &      & 44.50 & 18000 & 24 &  3 & 15400 \\ 
                 & 55277 & Yes &      & 44.50 & 18000 & 25 &  1 & 15000 \\ 
                 & 55287 & No &      & 44.35 & 16500 & 33 &  $<$ 50 & 12400 \\ 
                 & 55294 & No &      & 44.25 & 15000 & 39 &  $<$ 1 & 11300 \\ 
                 & 55295 & No &      & 44.20 & 14750 & 40 &  $<$ 100 & 10900 \\ 
                 & 55308 & No &      & 44.00 & 12250 & 50 &  $<$ 300 & 9900 \\ 
                 & 55318 & No &      & 43.80 & 11500 & 59 &  $<$ 100000 & 8700 \\ 
SN2011kg & 55922 & Yes & 0.1 & 44.00 & 10500 & 27 &  1 & 14800 \\ 
                 & 55926 & Yes &      & 44.00 & 10000 & 30 &  1 & 14500 \\ 
                 & 55935 & Yes &      & 43.90 & 9250 & 37 &  10 & 12700 \\
                  & 55943 & No &       & 43.85 & 8500 & 44 &  $<$ 30 & 11600 \\ 
                  & 55944 & No &       & 43.80 & 8500 & 45 &  $<$ 300 & 10900 \\ 
                  & 55952 & No &       & 43.65 & 7500 & 52 &  $<$ 500 & 10200 \\
                  & 55958 & No &      & 43.55 & 7000 & 57 &  $<$ 1000 & 9600 \\ 
PTF12dam & 56067 & Yes & 0.4 & 44.20 & 10750 & 42 &  10 & 12700 \\ 
                 & 56068 & Yes &       & 44.30 & 10750 & 42 &  5 & 13700 \\ 
                 & 56069 & Yes &       & 44.25 & 10750 & 43 &  10 & 13100 \\
                 & 56070 & Yes &       & 44.25 & 10500 & 44 &  10 & 13000 \\ 
                 & 56071 & Yes &       & 44.25 & 10500 & 45 &  10 & 12900 \\
                 & 56072 & Yes &       & 44.25 & 10500 & 46 &  10 & 12700 \\
                 & 56073 & Yes &       & 44.25 & 10250 & 47 &  10 & 12900 \\
                 & 56079 & Yes &       & 44.25 & 10000 & 52 &  30 & 12100 \\
                 & 56086 & Yes &      & 44.35 & 9500 & 59 &  10 & 12600 \\
                 & 56092 & No &       & 44.35 & 9250 & 64 &  $<$ 30 & 12200 \\
                 & 56096 & No &       & 44.35 & 9000 & 68 &  $<$ 30 & 12000 \\
                 & 56099 & No &       & 44.30 & 8750 & 70 &  $<$ 50 & 11500 \\
                 & 56107 & No &       & 44.25 & 8250 & 78 &  $<$ 100 & 11000 \\
                 & 56114 & No &       & 44.15 & 7750 & 84 &  $<$ 300 & 10400 \\
                 & 56119 & No &      & 44.15 & 7500 & 88 &  $<$ 300 & 10400 \\
                 & 56120 & No &       & 44.15 & 7500 & 89 &  $<$ 300 & 10300 \\
                 & 56125 & No &      & 44.10 & 7250 & 94 &  $<$ 500 & 9900 \\
                 & 56148 & No &     & 43.80 & 6250 & 115 &  $<$ 50000 & 8500 \\ 
iPTF13ajg & 56390 & Yes & 1.0 & 44.50 & 12250 & 49 &  30 & 13100 \\
                & 56391 & Yes &      & 44.50 & 12250 & 50 &  30 & 12900 \\
                & 56399 & Yes &      & 44.55 & 12250 & 54 &  50 & 12500 \\
                & 56420 & No &       & 44.55 & 11250 & 66 &  $<$ 100 & 11800 \\
                & 56422 & No &       & 44.55 & 11000 & 68 &  $<$ 100 & 11700 \\
                & 56449 & No &       & 44.20 & 9000 & 83 &  $<$ 1000 & 9900 \\
                & 56453 & No &       & 44.20 & 9000 & 85 &  $<$ 3000 & 9900 \\
                & 56485 & No &       & 43.95 & 8000 & 104 &  $<$ 100000 & 8400 \\ 
LSQ14mo & 56688 & Yes & 0.2 & 44.20 & 10500 & 33 &  1 & 14900 \\
                & 56694 & Yes &      & 44.20 & 10000 & 38 &  1 & 14100 \\
                & 56716 & No &       & 43.75 & 7750 & 56 &  $<$ 300 & 10300 \\
                & 56724 & No &       & 43.50 & 7250 & 62 &  $<$ 10000 & 8800 \\ 
SN2015bn & 57070 & No & 0.4 & 44.25 & 9000 & 63 &  $<$ 30 & 11900 \\
                 & 57071 & No &      & 44.25 & 9000 & 64 &  $<$ 30 & 11600 \\
                 & 57077 & No &      & 44.30 & 8750 & 69 &  $<$ 30 & 11800 \\
                 & 57078 & No &      & 44.30 & 8750 & 70 &  $<$ 30 & 11500 \\
                 & 57082 & No &      & 44.35 & 8750 & 74 &  $<$ 30 & 11500 \\
                 & 57092 & No &      & 44.35 & 8250 & 83 &  $<$ 50 & 11200 \\
                 & 57093 & No &      & 44.40 & 8250 & 84 &  $<$ 50 & 11600 \\
                 & 57099 & No &      & 44.40 & 8000 & 89 &  $<$ 50 & 11600 \\
                 & 57105 & No &      & 44.40 & 7750 & 94 &  $<$ 50 & 11400 \\
                 & 57108 & No &      & 44.40 & 7500 & 97 &  $<$ 50 & 11500 \\
                 & 57123 & No &      & 44.25 & 7000 & 110 &  $<$ 500 & 10200 \\
                 & 57135 & No &      & 44.05 & 6500 & 121 &  $<$ 3000 & 9100 \\
                 & 57136 & No &      & 44.05 & 6500 & 122 &  $<$ 3000 & 9000 \\
                 & 57150 & No &      & 44.00 & 6000 & 135 &  $<$ 10000 & 8900
\enddata
\label{table:param}
\end{deluxetable*}

\end{document}